\documentclass[aps,pre,reprint,superscriptaddress,showkeys,longbibliography,nofootinbib,floatfix]{revtex4-2}

\usepackage{amsmath,amssymb}
\usepackage{graphicx}
\usepackage{booktabs}
\usepackage[colorlinks=true,allcolors=blue]{hyperref}

\newcommand{\authornote}{All authors contributed equally to
this work and are listed alphabetically.
Correspondence may be addressed to
\href{mailto:igoshin@rice.edu}{igoshin@rice.edu},
\href{mailto:makarov@cm.utexas.edu}{makarov@cm.utexas.edu}, or
\href{mailto:morozov@physics.rutgers.edu}{morozov@physics.rutgers.edu}.}

\begin{document}

\title{Universal power law and condensation of wealth or population in nonequilibrium exchange games}

\author{Oleg A. Igoshin}
\thanks{\authornote}
\affiliation{Center for Theoretical Biological Physics, Rice University,
Houston, TX 77005}
\affiliation{Department of Chemistry, Rice University, Houston, TX 77005}
\affiliation{Department of Bioengineering, Rice University, Houston, TX 77005}
\affiliation{Department of Biosciences, Rice University, Houston, TX 77005}

\author{Dmitrii E. Makarov}
\thanks{\authornote}
\affiliation{Department of Chemistry, University of Texas at Austin,
Austin, TX 78712}
\affiliation{Oden Institute for Computational Engineering and Sciences,
University of Texas at Austin, Austin, TX 78712}

\author{Alexandre V. Morozov}
\thanks{\authornote}
\affiliation{Department of Physics and Astronomy, Rutgers University, Piscataway, NJ 08854-8019}


\makeatletter
\def\frontmatter@preabstractspace{24pt}
\def\@keys@name{\textbf{Keywords:} }
\makeatother

\begin{abstract}
\noindent
Power-law distributions characterize many complex systems, including wealth and city-size distributions, yet their origin and limits remain unclear. Why do similar scaling laws emerge from different underlying dynamics, and how can exceptionally large outliers arise beyond the apparent power-law background? Here we show that both phenomena can emerge in a deliberately minimal nonequilibrium exchange model. Agents exchange a conserved quantity in discrete units, with the direction of transfer biased only by the difference between their current holdings. Across a broad class of exchange rules biased in favor of the richer agent, the long-time distribution develops the same Pareto tail, \(p(m)\sim m^{-2}\), independent of the detailed form and strength of the bias. Although this scale-free distribution has a divergent mean in the infinite-system limit, finite populations can sustain a truncated version of it up to a
size-dependent wealth capacity. Beyond this capacity, additional wealth leaves the distribution of the great majority of agents nearly unchanged and instead condenses onto a single dynamically selected agent, producing an exceptionally large outlier, or
``dragon king.'' Consistent with these predictions, an analysis of city populations across 53 countries reveals a distribution close to the predicted Pareto law and an excess of unusually dominant largest cities. Thus, our results establish a unified mechanism for universal scaling and wealth/population condensation, and illustrate the coexistence of their statistical signatures in empirical data.

\medskip
\noindent\textbf{Significance.}
Wealth and city populations often exhibit power-law distributions, but why similar scaling laws emerge from different underlying processes remains unclear. We show that a broad class of simple nonequilibrium exchange models produces inverse square power-law distributions.
This universal behavior requires only exchanges favoring wealthier agents, but is independent of the detailed form of this bias. In finite populations, this distribution can coexist with a single exceptionally wealthy agent that accumulates most of the excess wealth. An analysis of city populations across 53 countries reveals both approximate Pareto scaling and an excess of unusually dominant largest cities. These findings provide a unified explanation for the emergence of universal power laws and exceptional outliers from simple exchange dynamics.
\end{abstract}


\maketitle

Econophysics models involving agents exchanging assets are useful both as models of wealth distribution in real-world economies~\cite{yakovenko_colloquium_2009,greenberg_twenty-five_2024} and as simple  models of statistical mechanics, where fundamental laws of statistical mechanics can be derived directly from dynamical laws rather than being imposed as postulates~\cite{miao_nonequilibrium_2024,makarov_what_2025}.  In particular, the canonical distribution arises naturally in such models as a consequence of time-reversal symmetry~\cite{yakovenko_colloquium_2009,lanchier_rigorous_2017,lanchier_rigorous_2018,cao_derivation_2023,cao_interacting_2024}. At the same time, the canonical distribution is widely applicable to wealth distributions, at least when the wealth of the richest subpopulations is not considered~\cite{yakovenko_colloquium_2009}.

Wealth distributions governing the wealthy, however, are often non-exponential: instead, they are usually well described by power laws, $p(m)\propto m^{-\mu}$, with an exponent often close to $2$ (the Pareto law).  The origin of this type of distribution is not a settled question. Many proposed explanations are based on preferential attachment models~\cite{krapivsky_scale-free_2008,krapivsky_connectivity_2000,barabasi_emergence_1999}. While plausible, such models produce non-universal values of the exponent $\mu$ that are determined by the model parameters.  At the same time, it is generally understood that non-Boltzmann statistics of wealth is linked to ``inequality'' and/or nonequilibrium character of the wealth dynamics; thus motivated, a number of econophysics studies have been undertaken  to explore such effects~\cite{ispolatov_wealth_1998,boghosian_economically_2022,liu_simulation_2021,cao_derivation_2023}.

Similar power-laws are also observed in the distributions  of sizes of cities, earthquakes, and forest fires~\cite{newman_power_2005}, scientific citations~\cite{peterson_nonuniversal_2010,merton_matthew_1968,reed_pareto_2001}, as well as in many other phenomena.
Among these, the distribution
of city sizes is especially relevant to the models
discussed here. It approximately follows Zipf's law,
$m_k\propto k^{-1}$, where $m_k$ is the population of
the $k$th-largest city, equivalent to a Pareto density
$p(m)\propto m^{-2}$~\cite{gabaix_zipfs_1999}.
Moreover, population changes
through migration between cities provide a natural
analogy to wealth exchanges between agents.
In this
context, many different explanations of this law have been proposed~\cite{gabaix_zipfs_1999}, often relating its origin to the population growth process. Thus, the power law is attributed to nonstationary growth rather than stationary exchange dynamics.   But the specific value $\mu \approx 2$ observed for many different countries remains unexplained.

A recent study~\cite{miao_nonequilibrium_2024} finds that a simple model of random wealth exchange -- a modified  version of the original Bennati-Dragulescu-Yakovenko (BDY) model~\cite{dragulescu_statistical_2000,bennati_il_1993,yakovenko_colloquium_2009} -- predicts the Pareto wealth distribution with $\mu=2$.  Unlike the BDY model, where the probabilities of giving and receiving money are both equal to $1/2$,  the modified model assumed unequal probabilities of giving and receiving, whose values depend on whether the giver/receiver is richer than the exchange partner.
Since the total wealth of the population is constant, as in the BDY case,  a steady state is attained in this model, but, unlike the BDY case, it is a nonequilibrium steady state breaking the time-reversal symmetry. The predicted exponent, $\mu = 2$, is independent of the exchange probabilities and is close to the empirically observed exponents, raising the question of whether the underlying model captures some features of the real world.

Here we show, using a more general class of models where the probability of giving or receiving money in an exchange is a function of  the difference between the exchange partners' wealth, that the exponent $\mu=2$ is universal for a broad class of models of this type.  Notably, the total wealth is conserved in these models, so the power law does not arise from
growth or explicitly nonstationary dynamics. In applications to economies or cities,
the stationary approximation may remain useful when wealth or population changes slowly
relative to the relaxation of the exchange dynamics.
We further show that, for populations of finite size, there is a critical value of the total wealth beyond which the wealth of most of the population obeys the Pareto power-law, while the excess wealth is absorbed by a single population member. We provide a theory of this ``dragon-king''~\cite{sornette_dragon-kings_2012} phenomenon and test it with Monte Carlo simulations.  We further show that the Pareto-law regime, the Boltzmann wealth distribution in the BDY  case, as well as several other predictions of the model, can all be understood within a unified view wherein the wealth distributions are described by a family of simple second-order differential equations for the probability $S(m)$ that the wealth of a population member is greater than or equal to $m$.  Finally, motivated by the correspondence between wealth exchange and migration between cities, we analyze city-size distributions across 53 countries. We find that near-Zipf scaling is accompanied by a systematic excess of unusually dominant largest cities relative to power-law sampling expectations, illustrating the empirical relevance of the two statistical features predicted by the model.

\section*{The model}
\begin{figure}
    \centering
    \includegraphics[width=\linewidth]{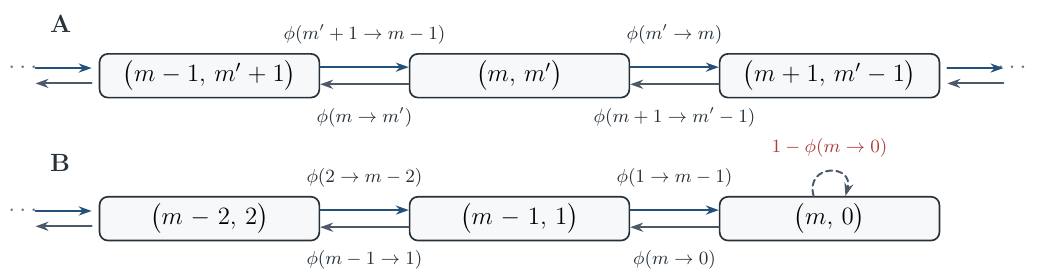}
    \caption{Wealth exchange game. (\textbf{A}) A local segment of the state network involving two agents with positive wealth. Arrows indicate allowed single-unit transfers between neighboring configurations. (\textbf{B}) A local segment of the state network adjoining a state where one of the agents has zero wealth. At $(m,0)$, a transfer from the wealthier agent to the empty agent occurs with probability $\phi(m\to 0)$, while an attempted transfer from the empty agent is rejected, leaving the state unchanged with probability $1-\phi(m\to 0)$. Here, $\phi(a\to b)=1-\phi(b\to a)$ denotes the probability that, in a pairwise interaction, one unit of wealth is transferred from an agent of wealth $a$ to an agent of wealth $b$.}
    \label{fig:kinetic scheme}
\end{figure}

Our model consists of $N\gg 1$ agents (players) exchanging money probabilistically (Fig.~\ref{fig:kinetic scheme}). The same model can be applied to describe random migration between $N$ cities, in which case agents are replaced by cities and a city's population is the equivalent of the wealth of an agent. At every step of the game, a pair of players is selected at random. Suppose that the wealth of the first player is $m$ and that of the second is $m'$, both being integers measured in units of money equal to the (fixed) amount exchanged in each transaction. In the case of inter-city migration, $m$ and $m'$ should be interpreted as their populations.  Assume, first, that $m,m'\ne 0$. Then, as a result of the exchange, the new amounts owned by the players are $m-1$ and $m'+1$ or $m+1$ and $m'-1$. The probability of the first outcome is $\phi(m\to m')$, with the arrow indicating the direction of wealth transfer, and the probability of the second is $\phi(m'\to m) = 1-\phi(m\to m')$. When $m=m'=0$, nothing happens as a result of the interaction between the two agents. If, say, $m>0$ but $m'=0$, the first player cannot receive money. In this case, we impose the following rule: the first player gives one unit of money with a probability $\phi(m\to m'=0)$; otherwise, no exchange takes place, with a probability $1-\phi(m\to 0)$.

We now focus on the evolution of wealth $m$ of a single player, which undergoes a random walk on the set of nonnegative integers. Every time the selected agent interacts with another agent, it steps left ($m \to m-1$), right ($m \to m+1$), or stays put ($m \to m$).  Within the mean-field theory, this random walk is a simple Markovian process governed by the transition probabilities for the steps~\cite{miao_nonequilibrium_2024}. Let $p(m+1|m)$ be the probability that the walker steps right. In the steady state, this is given, for any $m$, by
\begin{equation} \label{money up}
    p(m+1|m) =\sum_{m'=1}^{\infty} p(m')\phi(m'\to m)
\end{equation}
Here, $p(m')$ is the steady-state probability to have $m'$ units of wealth.
The sum runs from $m' = 1$ because players with $m'=0$ cannot give money.
We set the upper summation limit to infinity assuming that $p(m)$ decays quickly enough that the effect of finite population size can be neglected -- we will return to this issue later. Similarly, the probability $p(m-1|m)$ of losing one unit of money is
\begin{equation} \label{money down}
\begin{split}
        p(m-1|m) &=\sum_{m'=0}^{\infty} p(m')\phi(m\to m')
        \\
        &= \sum_{m'=0}^{\infty} p(m')[1-\phi(m'\to m)]
\end{split}
\end{equation}
for $m>0$. Since the wealth of a  non-empty player is unchanged only when the player encounters an empty player, the probability of having an unproductive encounter is
\begin{equation}
    p(m|m) =p(0) [1-\phi(m\to 0)] = p(0) \phi(0\to m),
\end{equation}
from which it follows that $p(m-1|m)+p(m|m)+p(m+1|m) = 1$ for $m>0$. For $m=0$, we have an obvious boundary condition $p(-1|0)=0$, preventing an agent that has nothing from losing more money. We also have
\begin{equation}
    p(0|0) = p(0) + \sum_{m'=1}^{\infty} p(m')[1-\phi(m'\to 0)].
\end{equation}
Eq.~\ref{money up} still holds for $m=0$, such that we have $p(0|0)+p(1|0)=1$.

As our random walk is one-dimensional, the steady-state probabilities satisfy detailed balance, which gives
\begin{equation} \label{DB}
\begin{split}
\frac{p(m+1)}{p(m)}&=\frac{p(m+1|m)}{p(m|m+1)}\\ &= \frac{\sum_{m'=1}^{\infty} p(m')\phi(m'\to m)}{\sum_{m'=0}^{\infty} p(m')[1-\phi(m'\to m+1)]}
\end{split}
\end{equation}
Suppose now that the probability $\phi(m\to m')$ depends only on the difference of $m$ and $m'$,
\[
\phi(m \to m')\equiv \phi(m-m') = 1-\phi(m'-m)
\]
Subtracting $1/2$ from the lhs and rhs of the above equation, we get
\[
\phi(m-m')-1/2 = 1/2-\phi(m'-m)\equiv -\eta(m-m'),
\]
where the new function $\eta(n)$ is odd, $\eta(-n)=-\eta(n)$. That is, we write
\begin{equation} \label{eta}
    \phi(n)=\frac{1}{2}-\eta(n).
\end{equation}
where $|\eta(n)|\le 1/2$. Substituting Eq.~\ref{eta} into Eq.~\ref{DB} and using $\sum_{m=0}^\infty p(m)=1$, we find
\begin{equation}\begin{split}
 \label{DB1}
\frac{p(m+1)}{p(m)}
&= \frac{\frac{1}{2}(1-p(0))-f(m)+p(0) \eta(-m)}{\frac{1}{2}+f(m+1)},
\end{split}
\end{equation}
where
\begin{equation} \label{f function}
 f(m)= \sum_{m'=0}^{\infty} p(m')\eta(m'-m).
\end{equation}

We will assume that $\eta(n)$ is a function that interpolates between $\lim_{n\to -\infty}\eta(n) \equiv -\eta_\infty $ and $\lim_{n\to \infty}\eta(n) \equiv \eta_\infty$. The value $\eta_\infty $ is positive for a ``rich-biased'' game, with the richer of the two interacting players being more likely to receive than give money. It is negative in the opposite case of a ``poor-biased'' game. The constant value $\eta(n)=0$ corresponds to the ``fair game'', in which the money exchange between two agents with non-zero wealth has the same probability in either direction. In this case one finds from Eq.~\ref{DB1}
\[
p(m)=p(0)[1-p(0)]^m,
\]
which is the Boltzmann distribution, where the value $p(0)$ is related to the average wealth per player, $\langle m \rangle = \sum_{m=0}^{\infty}mp(m)$ by
\begin{equation} \label{p0 Boltzmann}
    p(0)=\frac{1}{\langle m \rangle +1}.
\end{equation}
For the  general scenario, we can make progress by considering the asymptotic behavior of $p(m)$ at large $m$. In this case we have
\[f(m\to\infty)=-\eta_{\infty},\]
and substituting this result into Eq.~\ref{DB1}, we find
\begin{equation} \label{subcritical most general}
   \frac{p(m+1)}{p(m)} \xrightarrow[m \to \infty]{} \frac{\frac{1}{2}+\eta_{\infty}}{\frac{1}{2}-\eta_{\infty}} [1-p(0)].
\end{equation}
Provided that the expression on the right hand side of this equation is less than one, this shows that the distribution $p(m)$ will  have an exponential tail,
\begin{equation} \label{exponential tail}
p(m) \propto \left \{ \frac{\frac{1}{2}+\eta_{\infty}}{\frac{1}{2}-\eta_{\infty}} [1-p(0)] \right \}^m.
\end{equation}

\begin{figure}
    \centering
    \includegraphics[width=0.97\linewidth]{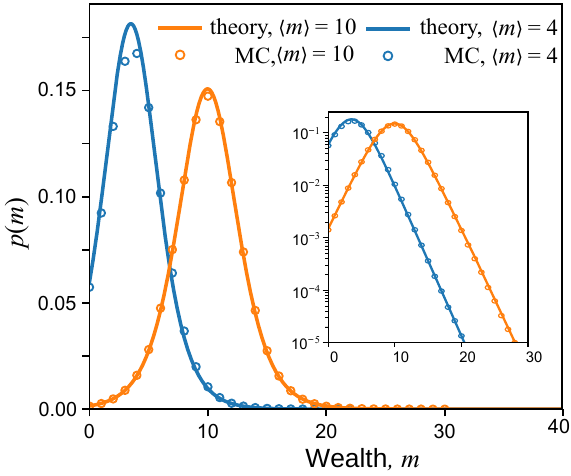}
    \caption{\textbf{Wealth distributions for poor-biased exchange.}
Continuum predictions (Eqs.~\ref{SI-theory poor biased} and
\ref{SI-mean wealth poor biased theory}) compared with Monte Carlo (MC)
simulations of $N=10{,}000$ agents at the indicated mean wealths.
The inset shows the same distributions on a logarithmic probability
scale, highlighting their exponential tails. Here, $|\eta_\infty| = 0.15$ (Table~\ref{SI-tab:SI_parameters}).}
    \label{fig:poor-biased-data}
\end{figure}

For the poor-biased game, where $\eta_\infty < 0$, the expression in Eq.~\ref{subcritical most general} is always less than $1$, and the exponential tail always exists. In this case, the wealth distribution typically shows a maximum, indicating a trend for wealth equalization (Fig.~\ref{fig:poor-biased-data}).  Treating $m$ as a continuous variable (which implies $ |\eta_\infty|\ll 1$) and additionally assuming   $p(0)\ll |\eta_\infty|$,  the distribution $p(m)$ is well approximated by the formula
\begin{equation} \label{especially simple}
    p(m)\approx \frac{|\eta_\infty|}{\cosh^2[2 |\eta_\infty|(m-\langle m\rangle)]},
\end{equation}
a function that is peaked at the mean value $\langle m\rangle$ and has an exponential tail described by Eq.~\ref{exponential tail}.  A more general analytical approximation for this case is derived in SI Appendix, Section~\ref{SI-sec:ContSol} and compared with Monte Carlo simulations of the game in Fig.~\ref{fig:poor-biased-data}.

The rich-biased game, where $0<\eta_\infty<1/2$, is more interesting.  If the right-hand side of Eq.~\ref{subcritical most general} were greater than 1, the probability $p(m)$ would grow indefinitely with $m$, which is impossible. Demanding that it does not exceed 1 results in an inequality
\begin{equation} \label{critical}
    p(0)\ge p_c = \frac{2\eta_\infty}{\eta_\infty + \frac{1}{2}},
\end{equation}
with a critical value $p_c$ where the $p({m+1})/p(m)$ approaches 1.

\begin{figure*}
    \centering
   \includegraphics[width=\linewidth]{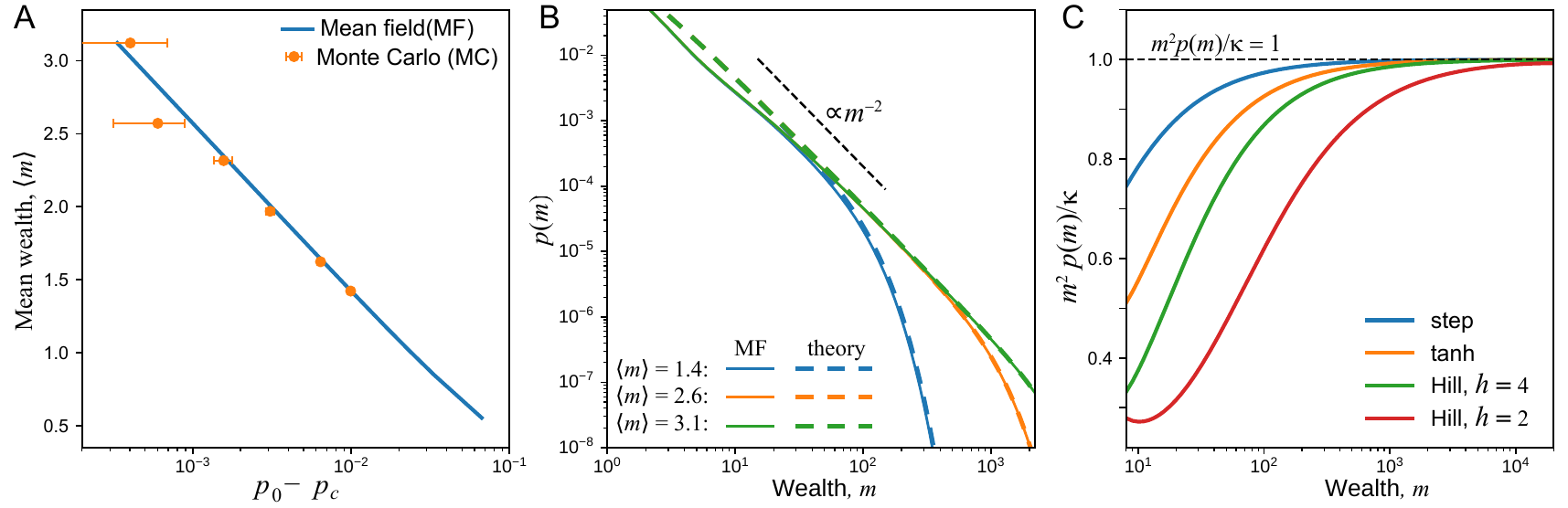}
   \caption{\textbf{Mean-field and Monte Carlo tests of Pareto scaling.}
Here $\eta_\infty=1/4$, $p_c=2/3$, and $\kappa=1/2$.
\textbf{(A)} Mean wealth versus $p(0)-p_c$ for
$\eta(n)=\eta_\infty\tanh(n/3)$, where $n$ denotes the wealth separation.
The solid line is the stationary mean-field (MF) prediction;
symbols show Monte Carlo results for $N=5{,}000$.
Horizontal error bars are standard errors across 12 independent runs.
Initial states were sampled from distributions calculated by
self-consistent iteration of Eqs.~\ref{DB1} and \ref{f function},
then equilibrated before averaging.
\textbf{(B)} Numerical MF distributions and continuum approximations
at $\langle m \rangle\simeq1.4,\,2.6,\,3.1$.
The continuum curves follow Eq.~\ref{SI-distribution subcritical},
with the matching constant $m_0$ fitted and $\delta$ and $\kappa$ fixed.
\textbf{(C)} Rescaled MF distributions $m^2p(m)/\kappa$
for step, tanh, and Hill functions $\eta(n)$, at
$\delta = 10^{-5}$ (cf. Eq.~\ref{distribution subcritical}).
The step function is $\eta(n)=\eta_\infty\operatorname{sgn}(n)$;
the Hill functions are
$\eta(n)=\eta_\infty\operatorname{sgn}(n)
|n|^h/(3^h+|n|^h)$, with $h=2,4$.
The curves approach a common near-critical Pareto regime over
the plotted range; the eventual subcritical cutoff lies beyond it.}
    \label{fig:Pareto}
\end{figure*}

Suppose now that the total wealth $M=N \langle m \rangle =N\sum_m{m}p(m)$ in the game is very small, say $M\ll N$. It is obvious, then, that $p(0)$ is very close to $1$. If $M$ is increased, the probability $p(0)$ will decrease. Figure~\ref{fig:Pareto}A illustrates this trend. According to Eq.~\ref{exponential tail}, this will result in slower exponential decay of the distribution tail. What happens when $p(0)$ approaches the critical value $p_c$ from above?  The existence of the exponential tail ensures that the average wealth $\langle m \rangle $ is finite for any $p(0)>p_c$, but it may diverge exactly in the critical case $p(0)=p_c$, where Eq.~\ref{exponential tail} no longer holds~\cite{miao_nonequilibrium_2024}.  This regime is the most interesting one and is the main focus of this paper.

\section*{The critical case and the universal Pareto law}

To analyze the wealth distribution in the critical case, we return to Eq.~\ref{DB1}. If the transition between the two limits of $\eta(n)$, $-\eta_{\infty}$ and $+\eta_{\infty}$, is ``sharp enough'', then $\eta(n)$ can be approximated by
\begin{equation}\label{step-function}
    \eta(n)\approx \eta_\infty [2\theta(n)-1],
\end{equation}
where $\theta(n)$ is the Heaviside step-function. In this case we obtain
\begin{equation} \label{key approx}
\begin{split}
    f(m) &\approx -\eta_\infty \sum_{m'=0}^{m-1}p(m') +  \eta_\infty \!\!\sum_{m'=m+1}^{\infty}p(m') \\
    &= -\eta_\infty [1-S(m)] + \eta_\infty S(m+1) \\
    &\approx \eta_{\infty} [2 S(m)-1],
    \end{split}
\end{equation}
where
\begin{equation} \label{tail}
    S(m)=\sum_{m'=m}^{\infty} p(m')
\end{equation}
is the ``tail probability''. In the second equality, we used the fact that $S(m)\gg p(m)$, since this sum contains many terms comparable with $p(m)$.
Although Eq.~\ref{key approx} is, in general, approximate, it turns out to be {\it exact}  in the critical case assuming the steady state of Eq.~\ref{DB1} and  the limit $m\to\infty$, as long as $\eta(n)$ approaches its $n\to \infty$ limit {\it faster than the inverse power law} (SI Appendix,  Section~\ref{SI-sec:asymp_just}). This is a crucial observation, and it is the reason why the tail behavior  of the distribution $p(m)$ turns out to be universal for a broad class of models, regardless of the precise function $\eta(n)$.

Combining Eqs.~\ref{DB1} with Eq.~\ref{key approx}, we have
\begin{equation}\label{interation general}
    \frac{p(m+1)}{p(m)} \approx \frac{(\frac{1}{2}+\eta_{\infty})[1-p(0)]-2\eta_\infty S(m)}{\frac{1}{2}-\eta_{\infty}+2\eta_\infty S(m)}
\end{equation}
For $p(0)> p_c$ and $m\to \infty$, $S(m)$ is negligible, and the exponentially decaying probability,  Eq.~\ref{exponential tail},  is recovered. But in the critical regime, $p(0)=p_c$, $S(m)$ cannot be neglected. In this case we rewrite Eq.~\ref{interation general} as follows:
\begin{equation} \label{interation critical}
\frac{p(m+1)}{p(m)}\approx \frac{\frac{1}{2}-\eta_{\infty}-2\eta_\infty S(m)}{\frac{1}{2}-\eta_{\infty}+2\eta_\infty S(m)}
\approx 1 - \frac{2}{\kappa} S(m),
\end{equation}
where we have defined $\kappa = ({\frac{1}{2}-\eta_{\infty}})/{2\eta_\infty}$.

Approximating $m$ as a continuous variable, with $dp(m)/dm \approx p(m+1)-p(m)$, we rewrite this as a differential equation,
\begin{equation} \label{diff E for distribution}
\frac{dp}{dm}= -\frac{2}{\kappa} p(m) S(m),
\end{equation}
which, after recognizing the relationship
\begin{equation} \label{relation between S and p}
    p(m)=-\frac{dS(m)}{dm},
\end{equation}
becomes a second-order differential equation
\begin{equation}
-\frac{d^2S}{dm^2}= \frac{2}{\kappa} \frac{dS}{dm} S = \frac{1}{\kappa} \frac{d(S^2)}{dm}.
\end{equation}
This equation is easily integrated; noting that $S\to 0$ for $m\to\infty$, one finds
\begin{equation} \label{Pareto}
   S(m)\approx  \frac{\kappa}{m} ~\text{ and }~ p(m)\approx  \frac{\kappa}{m^2}.
\end{equation}
Eq.~\ref{Pareto} is a power (Pareto)  law with a universal scaling exponent of $-2$.  The distribution is normalized, but its first moment $\langle m \rangle$ diverges: the critical regime corresponds to infinite wealth per player. Yet, unlike the case of the Boltzmann distribution (cf. Eq.~\ref{p0 Boltzmann}), the probability $p(0)$ of having zero wealth remains finite in this case.

Figures~\ref{fig:Pareto}A,B illustrate how the critical regime is approached as $p(0)$ approaches $p_c$ from above and the mean wealth diverges. Figure \ref{fig:Pareto}A also illustrates the validity of the mean-field approximation by comparing it with numerical Monte Carlo simulations.
Figure~\ref{fig:Pareto}B  further shows that the numerically computed mean-field distributions
develop an increasingly broad $m^{-2}$ regime as $p(0)$ approaches
$p_c$. For any \(p(0)>p_c\), however, the distribution ultimately crosses over to an exponential tail, as predicted by Eq.~\ref{exponential tail} (Fig.~\ref{fig:Pareto}B). In fact, sufficiently close to the critical regime, a more general expression that interpolates between  Eq.~\ref{Pareto} and the exponential tail of Eq.~\ref{exponential tail} is derived in SI Appendix, Section~\ref{SI-sec:NearCrit}:
\begin{equation} \label{distribution subcritical}
    p(m) \approx \frac{\delta^2 \kappa}{4\sinh^2[\delta (m-m_0)/2]},
\end{equation}
where $\delta = (p(0)-p_c)/(1-p_c)$
and $m_0$ is a constant. As seen from Fig.~\ref{fig:Pareto}B, this equation, indeed, describes the crossover between the  power-law and exponential regimes. Finally, Fig.~\ref{fig:Pareto}C illustrates that $p(m)$ approaches the Pareto power law, Eq.~\ref{Pareto}, as $m\to\infty$, regardless of the shape of the function $\eta(n)$, with the rate of convergence depending on how rapidly $\eta(n)$ approaches its asymptotic value $\eta_\infty$.

\section*{Finite-size effects: wealth condensation and the dragon kings }

\begin{figure*}
\centering
\includegraphics[width=\linewidth]{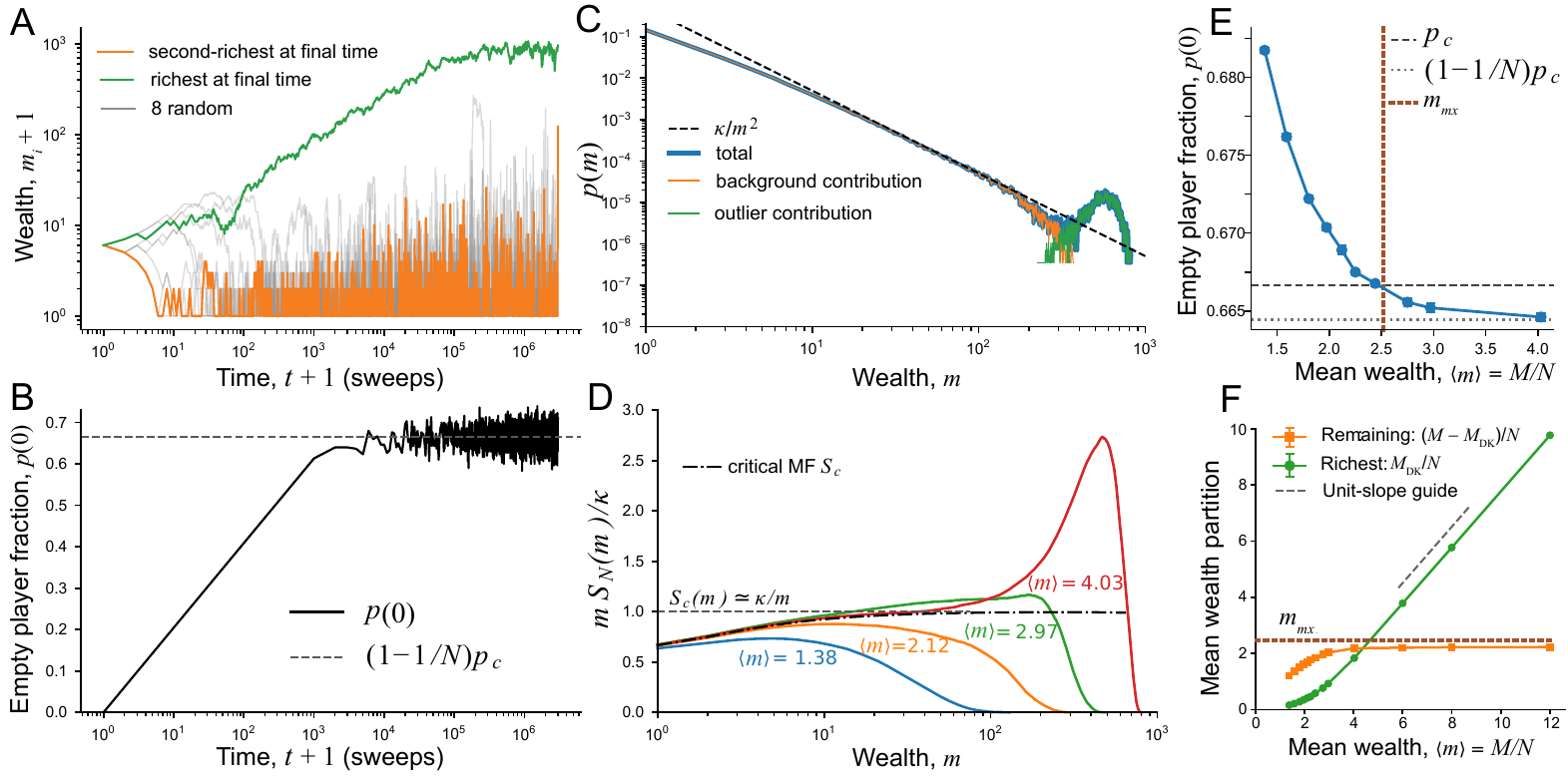}

\caption{\textbf{Wealth condensation in finite populations.}
All panels use $N=300$.
\textbf{(A)} Emergence from equal wealth $m_i=5$ ($M=1{,}500$);
highlighted trajectories follow fixed agents selected by their final ranks.
\textbf{(B)} Fraction of agents with zero wealth along the same trajectory, compared with
$(1-1/N)p_c$, the theoretically predicted fraction of zero-wealth agents in equilibrium with a single outlier.
\textbf{(C)} Stationary distribution at $M=1{,}209$, obtained by sampling from long MC trajectories at multiple time points
and averaged over four independent runs. The background distribution is obtained by discarding the wealthiest individual
at each time point; the background distribution weight is $1-1/N$.
\textbf{(D)} Rescaled tail probability $m S_N(m)/\kappa$ at the indicated values of
mean wealth, compared with the critical mean-field (MF) reference.
\textbf{(E)} Fraction of agents with zero wealth versus $\langle m\rangle=M/N$.
The vertical line marks the crossover estimate
$m_{mx}\simeq\kappa\ln(\kappa N)\simeq2.505$ (Eq.~\ref{eq:crossover_matching}).
\textbf{(F)} Mean-wealth contributions from the richest agent,
$M_{DK}/N$, and the remaining population,
$(M-M_{DK})/N$.
At high mean wealth, the remaining population's contribution levels off,
while that of the richest agent grows with slope close to unity.
The dashed line is a unit-slope guide.
Error bars in E and F are standard errors across twelve independent
trajectory averages.}

\label{fig:finiteN-one-outlier}
\end{figure*}
For finite populations, the rich-biased model exhibits a remarkable phenomenon: when the total wealth \(M\) exceeds a population-size-dependent crossover, a single player typically accumulates wealth far exceeding that of the remaining population. The wealth of the remaining players is approximately described by the Pareto distribution (Eq.~\ref{Pareto}) followed by a finite-size cutoff. This condensation of excess wealth into a single dynamically selected agent (``a dragon king~\cite{sornette_dragon-kings_2012}''), is illustrated in Fig.~\ref{fig:finiteN-one-outlier}A for a population of $N=300$ players.
This effect can be viewed as a condensation phenomenon analogous to those arising in various branches of physics, from Bose-Einstein condensation to glassy dynamics~\cite{ritort_glassy_2003,bialas_condensation_1997}. It was also predicted in other models of wealth dynamics~\cite{bouchaud_wealth_2000, burda_wealth_2002,boghosian_oligarchy_2017}.

To understand the dragon-king phenomenon, we first show that a single dragon-king can coexist with the rest of the population. Consider the wealthiest player in the population. Assuming Eq.~\ref{step-function}, the probabilities that the wealthiest player gains or loses one unit of wealth in the next encounter are, respectively,
\begin{align}
p_+&=
\frac{\left(\frac12+\eta_\infty\right)(N-1-Z)
}
{N-1},
\label{eq:app-Ui}\\
p_-&=
\frac12-\eta_\infty,
\label{eq:app-Di}
\end{align}
where $Z$ is the total number of players with zero wealth.  The first of these two probabilities,  $p_+$, decreases with increasing $Z$, since an encounter with a player that has no money cannot produce increase in wealth. If $Z$ is sufficiently large, the wealthiest player will tend to get poorer. Conversely, if $Z$ is small, the richest player will tend to get even richer. Thus, a single wealthy player reaches equilibrium with the rest of the population at a critical value of $Z$ for which $p_+=p_-$:
\begin{equation} \label{critical Z}
    Z=Z_c = p_c (N-1),
\end{equation}
with $p_c$ given by Eq.~\ref{critical}.  That is, a single wealthy outlier can coexist with the remaining $N-1$ players when the fraction of empty players within this remaining population is exactly at the critical value  $p_c$ (Fig.~\ref{fig:finiteN-one-outlier}B)!   This corresponds to $p(0)=Z/N = (1-1/N)p_c$, when the fraction of zero-wealth players is measured with respect to the entire population. It is also possible (statistically) to have more than one outlier accumulating significant wealth. Indeed, treating the top two wealthy players, collectively, as a combined super-player, we can repeat the above analysis to establish that they will not be exchanging, on the average, any wealth with the rest of the players when the number of players with zero wealth is $Z_c=p_c(N-2)$.  But consider the exchanges between these two players: whenever the amounts belonging to them are unequal, the exchange is a biased random walk with the probabilities  $1/2 +\eta_\infty$ and $1/2-\eta_\infty $  of, respectively, increasing and decreasing their wealth inequality. This biased competition drives coarsening toward a single dominant outlier, analogous to Ostwald ripening.

Now consider the rest of the population excluding the dragon-king(s). Starting with $p(0)=p_c$ and applying  Eq.~\ref{interation critical} to the subpopulation not containing the richest player, one would conclude that it should be  described by the Pareto law, Eq.~\ref{Pareto}. But this power law cannot hold indefinitely, since $m$ cannot exceed the total wealth $M$. In fact, in Figs.~\ref{fig:finiteN-one-outlier}C,D we observe a crossover $m\sim m^*$ separating Pareto-like behavior from a dragon-king peak observed at higher values of $m$.  To understand the origin of this crossover and to estimate $m^*$, we note that using Eq.~\ref{interation critical} here neglects the infrequent encounters of the subpopulation members with the dragon king, an effect of order $1/(N-1)$. While small for large $N$, these encounters modify the transition probabilities $p(m\pm1|m)$, resulting in an exponential cutoff on a wealth scale of order $\kappa N$ (SI Appendix, Section~\ref{SI-sec:finite_bulk}).  We thus estimate
\begin{equation}
m^* \simeq \kappa N.
\end{equation}
Using this value, we can further estimate the maximum wealth that can be absorbed by the ``homogeneous'' part of the population  as
\begin{equation}\label{eq:crossover_matching}
\begin{split}
M_{mx}&= (N-1)m_{mx} \ \text{with\ }\\
m_{mx}&\approx\sum_{m=1}^{m^*}m\,\frac{\kappa}{m^2} \approx \kappa \ln (\kappa N)
\end{split}
\end{equation}
In other words, the average wealth of the players excluding the outlier cannot exceed $m_{mx}\simeq \kappa \ln N +O(1)$ as $N\to\infty$, while the wealth of a single outlier can be much greater than that.

For a sufficiently small total wealth $M<M_{mx}$, the finite-size effects are unimportant and the mean-field theory of the previous section, assuming $N\to\infty$, describes the wealth distribution well. Accordingly, as $M$ increases, the zero-wealth fraction $p(0)$ approaches $p_c$ from above. However, for finite and fixed $N$, instead of attaining this value for $M\to\infty$, $p(0)$ drops below $p_c$ and continues to decrease, approaching a smaller value $p(0)\to p_c(1-1/N)$, the critical limit for the $N-1$ players which excludes the outlier (Fig. \ref{fig:finiteN-one-outlier}E). It is plausible that, at the onset of wealth condensation, where $M \approx M_{mx}$ (i.e., just before the outlier emerges), one has $p(0) \approx p_c$, and indeed this is supported by simulations (Fig.~\ref{fig:finiteN-one-outlier}E).

If the total wealth $M$ significantly exceeds $M_{mx}$, the excess wealth accumulates in the dragon king, while the mean wealth of the remaining population approaches a constant
\begin{equation}
    M_{DK} \approx M-M_{mx},
\end{equation}
The dragon-king wealth, \(M_{\rm DK} \equiv \langle \max_i (m_i) \rangle\), thus depends linearly on $M$,
a dependence verified numerically in Fig.~\ref{fig:finiteN-one-outlier}F.

\section*{Example: Pareto law and extreme outliers in city-size distributions}
\label{sec:city-results}
The exchange model has a simple interpretation in terms of migration between cities. Cities play the role of agents, population replaces wealth, and each move transfers one person from one city to another. A preference for larger destinations would correspond to rich-biased exchange. This idealized interpretation motivates asking whether city systems exhibit the two statistical features found above: a near-power-law background and an unusually dominant largest member.

City sizes have long been described by Zipf's rank--size rule, with the population of the $k$-th largest city scaling as $m_k\propto k^{-1}$~\cite{gabaix_zipfs_1999}. This rule is equivalent to a probability distribution of the city size given by
$p(m)\propto m^{-2}$, i.e., corresponding to the universal exponent of our simple model. Previous cross-country studies document
variation around this rule and sensitivity to city definitions and sample selection~\cite{RosenResnick1980,Soo2005}; exceptionally large leading cities
have also been investigated using statistical dragon-king tests~\cite{pisarenko_robust_2012}.

\begin{figure}[t]
\centering
\includegraphics[width=\columnwidth]{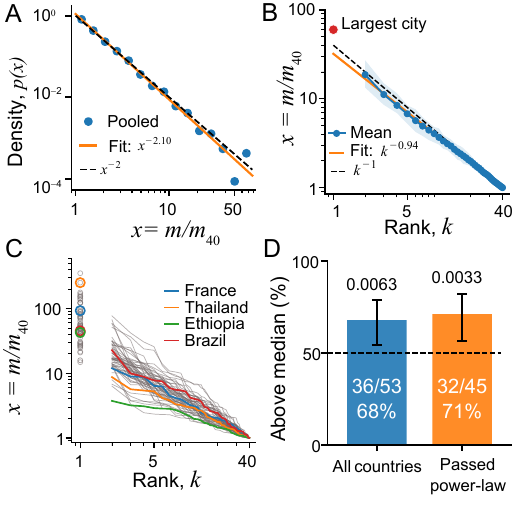}
\caption{\textbf{City-size scaling and leading-city excess.}
Populations are scaled by each country's 40th-largest city, $x=m/m_{40}$.
\textbf{(A)} Pooled density with the largest cities held out.
\textbf{(B)} Geometric mean across countries at each rank. Shading shows
the middle 50\% of country values, and the red point marks rank~1.
In A and B, orange lines show the background fits and dashed lines
show the Zipf references.
\textbf{(C)} Individual country profiles, with separate points for
the largest cities.
\textbf{(D)} Share of countries whose leading-city gap exceeds its
reference median under Pareto sampling, shown for all countries and
for those passing the power-law check. The dashed line marks the expected
50\%. Error bars show 95\% confidence intervals, and numbers above the bars
are one-sided binomial $p$-values.}
\label{fig:city-sizes}
\end{figure}

To assess this systematically with the most recent data, we started with the United Nations 2025 dataset~\cite{UNWUP2025} and selected the countries with at least 40 urban centers populated by at least 50,000 inhabitants.  Using the same 40 ranks gives each of the 53 countries an equal number of data points and
focuses the comparison on major cities. Reducing the minimum number of qualifying cities per country from $K=40$ to $K=30$ or $20$ expands the sample from 53 to 61 or 79 countries, respectively, while preserving the main findings reported below (SI Appendix, Section~\ref{SI-sec:CityMethods}). For each country, we divide each city's population by that of the 40th-largest one, $x=m/m_{40}$, and compute the probability distribution for the normalized city sizes across the countries. With each country's largest city left out, the
power-law fit to the combined data has an exponent of $-2.10$, close to
the Zipf value of $-2$ (Fig.~\ref{fig:city-sizes}A). At each rank $k$, we take the geometric mean of these normalized populations across countries. Fitting a power law over ranks 2--40  gives the scaling of $k^{-0.94}$, close to $k^{-1}$ in the Zipf law (Fig.~\ref{fig:city-sizes}B).

The geometric mean for the largest city size, however,  lies above the Zipf law trend. This could be viewed as evidence that the largest cities are dragon kings. However, they could also be merely ``black swans''; that is, sampling from the long tail of a power law can put the largest city well
above the rank trend.
Therefore, to test whether the leading cities are dragon-kings, we ask whether the gap between the two largest cities is unusually large given the gaps among the other major cities in the same country.

Figure~\ref{fig:city-sizes}C shows the
individual country profiles, with four examples highlighted. The slopes
vary across countries, as does the gap between the two largest cities. To see how far the largest city stands apart within its own country, we compare the gap between the two largest cities with the gaps among the other
major centers, using the Pareto benchmark of the statistical Dragon King test from Ref.~\cite{pisarenko_robust_2012}.
Using the gaps among ranks 2--40, we calculate how large the top city would
be if its gap above the second-largest matched the median benchmark
(SI Methods). This gives reference populations of 2.25 million for Paris,
1.06 million for Bangkok, and 0.834 million for Addis Ababa. Their observed
populations are 4.2, 17.2, and 8.0 times those values, respectively. In contrast,
S\~ao Paulo is about 4\% below its reference of 19.75 million
(Fig.~\ref{fig:city-sizes}C).

If all 40 city sizes in each country were independent draws from that country’s power law, we would expect about half of the largest cities to exceed their reference populations. Instead, 36 of 53 countries
(68\%) do so; using the approaches outlined in SI Appendix, Section \ref{SI-sec:CityMethods},  we compute the probability of that happening by chance under the power-law null hypothesis as $p=0.0063$
(Fig.~\ref{fig:city-sizes}D). Restricting the analysis to the 45 countries that are well-described by power-law gives 32 of 45 (71\%; $p=0.0033$). These observations provide an empirical example of near-Zipf pooled scaling
accompanied by a collective excess of leading-city dominance
consistent with the two statistical features produced by our model.

\section*{Conclusions}
Econophysics models with asset-exchanging agents have produced many insights both about wealth distributions~\cite{dragulescu_exponential_2001,dragulescu_statistical_2000,yakovenko_colloquium_2009,cao_mean-field_2025,blom_hallmarks_2025,boghosian_economically_2022,boghosian_fokkerplanck_2014,boghosian_economically_2022,sinha_stochastic_2003,ispolatov_wealth_1998,aydiner_universal_2023} and about foundational issues in statistical mechanics~\cite{miao_nonequilibrium_2024,lucente_h-theorem_2025, lanchier_rigorous_2017,lanchier_rigorous_2018,lanchier_distribution_2022}. Here, we studied a simple agent-based wealth exchange model, in which the probability for an agent to give/receive money to/from another agent is a function of the difference in their wealth. The same model can also be used to describe population migration between cities, with a city being an ``agent'' and its population being the agent's wealth. Our study makes two central predictions: First, for a broad class of models,  with the exchange probability varying ``not too slowly'' as a function of the wealth disparity, we find that, in the limit of infinite population, $N\to\infty$, and infinite average wealth, $M/N\to\infty$, the wealth distribution obeys Pareto's power law, $p(m)\propto m^{-2}$ with a universal scaling exponent that is independent of the model parameters.  Although we do not claim that our model describes realistic economies, the similarity of the exponent to those observed in real-world scenarios is notable. For example, a fit to the pooled city-size background gives a density exponent of $-2.1$ (Fig. 5A).

Second, when the population $N$ is finite, we observe condensation of wealth when the total population wealth $M$ exceeds a critical value: in this case a (typically) single population member (``dragon king''~\cite{sornette_dragon-kings_2012}) coexists with the rest of the population;  the wealth of this remaining population is described by the same Pareto law with a power law of $m^{-2}$ over an intermediate range of wealth, followed by a finite-size cutoff.

This second finding is especially interesting. Large outliers have been observed in many settings from distributions of weather events to city populations~\cite{sinha_evidence_2006,sornette_dragon-kings_2012,coelho_double_2008}. This can be illustrated by the population of Paris, which far exceeds that of  any other city in France. But this example also illustrates the difficulty of differentiating dragon kings from ``black swans'', or single data points sampled from heavy-tailed distributions.  And although statistical tests have been proposed to differentiate dragon kings from black swans~\cite{sornette_dragon-kings_2012,pisarenko_robust_2012}, theoretical models~\cite{bouchaud_wealth_2000,yukalov_statistical_2012,liu_simulation_2021,boghosian_oligarchy_2017,burda_wealth_2002}, which are not  limited by data statistics, are especially valuable in studying the wealth condensation phenomenon.  Our model shows that universal power-law distributions and exceptional outliers need not have separate origins: both can arise naturally from the same population- or wealth- conserving nonequilibrium dynamics. These findings are consistent with a further examination of city populations across 53 countries.

\section*{Materials and Methods}
\paragraph*{Mean-field calculations}
We solved the discrete stationary equations self-consistently
on $m=0,\ldots,L$. The one-unit transition probabilities
were recomputed from the current distribution, and the
zero-flux recurrence (Eq.~5) was iterated with damping.
At fixed $p(0)$, the positive-wealth distribution was
normalized to $1-p(0)$. An outer bisection adjusted $p(0)$
when a prescribed mean wealth was required.
For Fig.~3B, the continuum approximation was matched to
the numerical MF distributions by fitting only the offset
$m_0$, minimizing squared differences between logarithmic
densities. Numerical settings and convergence checks are
given in the SI Appendix, Section~\ref{SI-sec:Num}.
\paragraph*{Monte Carlo simulations}
Agents carried nonnegative integer wealth with fixed total
$M$ and mean $\langle m\rangle=M/N$. Each sweep randomly
permuted the agents into $N/2$ disjoint pairs.
A unit transfer from $j$ to $i$ was proposed with
probability $1/2+\eta(m_i-m_j)$; otherwise the reverse
transfer was proposed. Transfers from empty donors were
rejected but counted toward elapsed time.
For Fig.~3A, populations were initialized by stratified
sampling from the MF distribution, with total wealth
adjusted to $M$.

Stationary observables were time-averaged within
trajectories; ensemble means and standard errors were
calculated across independent trajectory averages rather
than individual snapshots. Initialization procedures,
sampling schedules, and convergence and conservation
checks are detailed in the SI Appendix,
Section~\ref{SI-sec:Num}.

\paragraph*{City populations and statistics}
Within each country, we computed rank-adjusted logarithmic
spacings $W_{j}=j\ln(m_j/m_{j+1})$.
For the largest $K=40$ cities, the dragon-king (DK)
statistic $R=38W_1/\sum_{j=2}^{39}W_{j}$ follows an
$F_{2,76}$ distribution under independent Pareto sampling,
independently of the Pareto exponent~\cite{pisarenko_robust_2012}.
We assessed the excess of dominant largest cities using
a one-sided binomial test, assuming independence across
countries, and report 95\% Wilson score intervals.
Data selection, background power-law screening
($p_{\mathrm{fit}}\geq0.10$), simulation procedures,
and robustness analyses are described in the SI Appendix,
Section~\ref{SI-sec:CityMethods}.

\section*{Data, Materials, and Software Availability}
The Python code and simulation parameters required to reproduce Figs.~2--5 and the numerical checks in the SI Appendix are publicly available at \url{https://github.com/OlegIgoshin/Universal-Pareto-Dragon-Kings}. The city-population data are from Ref.~\cite{UNWUP2025}.

\section*{Author Declaration}
The authors declare no competing interests.

\begin{acknowledgments}
The authors are grateful to the Telluride Science and Innovation Center for hosting a workgroup where this work was initiated.
\end{acknowledgments}

\bibliography{references}

@article{yakovenko_colloquium_2009,
  title     = {Colloquium: {Statistical} mechanics of money, wealth, and income},
  volume    = {81},
  url       = {https://doi.org/10.1103/RevModPhys.81.1703},
  doi       = {10.1103/RevModPhys.81.1703},
  number    = {4},
  journal   = {Rev. Mod. Phys.},
  publisher = {American Physical Society},
  author    = {Yakovenko, Victor M. and Rosser, J. Barkley},
  month     = dec,
  year      = {2009},
  pages     = {1703--1725}
}

@article{greenberg_twenty-five_2024,
  title   = {Twenty-five years of random asset exchange modeling},
  volume  = {97},
  issn    = {1434-6036},
  doi     = {10.1140/epjb/s10051-024-00695-3},
  number  = {6},
  journal = {The European Physical Journal B},
  author  = {Greenberg, Max and Gao, H. Oliver},
  month   = jun,
  year    = {2024},
  pages   = {69}
}

@article{miao_nonequilibrium_2024,
  title     = {Nonequilibrium statistical mechanics of money/energy exchange models},
  volume    = {57},
  url       = {https://dx.doi.org/10.1088/1751-8121/ad369b},
  doi       = {10.1088/1751-8121/ad369b},
  number    = {15},
  journal   = {J. Phys. A},
  publisher = {IOP Publishing},
  author    = {Miao, Maggie and Makarov, Dmitrii E. and Blom, Kristian},
  month     = apr,
  year      = {2024},
  pages     = {155003}
}

@article{makarov_what_2025,
  title   = {What the {Boltzmann} {Money} {Game} {Teaches} {Us} about {Statistical} {Mechanics} (and {Maybe} {Economics})},
  volume  = {16},
  issn    = {1948-7185, 1948-7185},
  url     = {https://pubs.acs.org/doi/10.1021/acs.jpclett.5c02426},
  doi     = {10.1021/acs.jpclett.5c02426},
  number  = {38},
  journal = {The Journal of Physical Chemistry Letters},
  author  = {Makarov, Dmitrii E.},
  month   = sep,
  year    = {2025},
  pages   = {9894--9899}
}

@article{lanchier_rigorous_2017,
  title   = {Rigorous {Proof} of the {Boltzmann}–{Gibbs} {Distribution} of {Money} on {Connected} {Graphs}},
  volume  = {167},
  issn    = {1572-9613},
  doi     = {10.1007/s10955-017-1744-8},
  number  = {1},
  journal = {Journal of Statistical Physics},
  author  = {Lanchier, Nicolas},
  month   = apr,
  year    = {2017},
  pages   = {160--172}
}

@article{lanchier_rigorous_2018,
  title   = {Rigorous {Results} for the {Distribution} of {Money} on {Connected} {Graphs}},
  volume  = {171},
  issn    = {1572-9613},
  doi     = {10.1007/s10955-018-2024-y},
  number  = {4},
  journal = {Journal of Statistical Physics},
  author  = {Lanchier, Nicolas and Reed, Stephanie},
  month   = may,
  year    = {2018},
  pages   = {727--743}
}

@article{cao_derivation_2023,
  title   = {Derivation of wealth distributions from biased exchange of money},
  volume  = {16},
  issn    = {1937-5093},
  doi     = {10.3934/krm.2023007},
  number  = {5},
  journal = {Kinetic and Related Models},
  author  = {Cao, Fei and Motsch, Sebastien},
  month   = oct,
  year    = {2023},
  pages   = {764--794}
}

@article{cao_interacting_2024,
  title     = {From interacting agents to {Boltzmann}-{Gibbs} distribution of money},
  volume    = {37},
  issn      = {0951-7715},
  doi       = {10.1088/1361-6544/ad8c8c},
  number    = {12},
  journal   = {Nonlinearity},
  publisher = {IOP Publishing},
  author    = {Cao, Fei and Jabin, Pierre-Emmanuel},
  month     = nov,
  year      = {2024},
  pages     = {125020}
}

@article{krapivsky_scale-free_2008,
  title   = {Scale-free networks as preasymptotic regimes of superlinear preferential attachment},
  volume  = {78},
  issn    = {1539-3755, 1550-2376},
  url     = {https://link.aps.org/doi/10.1103/PhysRevE.78.026114},
  doi     = {10.1103/PhysRevE.78.026114},
  number  = {2},
  journal = {Physical Review E},
  author  = {Krapivsky, Paul and Krioukov, Dmitri},
  month   = aug,
  year    = {2008},
  pages   = {026114}
}

@article{krapivsky_connectivity_2000,
  title   = {Connectivity of {Growing} {Random} {Networks}},
  volume  = {85},
  issn    = {0031-9007, 1079-7114},
  url     = {https://link.aps.org/doi/10.1103/PhysRevLett.85.4629},
  doi     = {10.1103/PhysRevLett.85.4629},
  number  = {21},
  journal = {Physical Review Letters},
  author  = {Krapivsky, P. L. and Redner, S. and Leyvraz, F.},
  month   = nov,
  year    = {2000},
  pages   = {4629--4632}
}

@article{barabasi_emergence_1999,
  title   = {Emergence of {Scaling} in {Random} {Networks}},
  volume  = {286},
  issn    = {0036-8075, 1095-9203},
  url     = {https://www.science.org/doi/10.1126/science.286.5439.509},
  doi     = {10.1126/science.286.5439.509},
  number  = {5439},
  journal = {Science},
  author  = {Barabási, Albert-László and Albert, Réka},
  month   = oct,
  year    = {1999},
  pages   = {509--512}
}

@article{ispolatov_wealth_1998,
  title     = {Wealth distributions in asset exchange models},
  volume    = {2},
  url       = {https://doi.org/10.1007/s100510050249},
  doi       = {10.1007/s100510050249},
  journal   = {The European Physical Journal B-Condensed Matter and Complex Systems},
  publisher = {Springer},
  author    = {Ispolatov, Slava and Krapivsky, Paul L and Redner, Sidney},
  year      = {1998},
  pages     = {267--276}
}

@article{boghosian_economically_2022,
  title   = {An economically realistic asset exchange model},
  volume  = {380},
  issn    = {1471-2962 (Electronic) 1364-503X (Linking)},
  doi     = {10.1098/rsta.2021.0167},
  number  = {2224},
  journal = {Philos Trans A Math Phys Eng Sci},
  author  = {Boghosian, B. M. and Hudes, M. and Khachatryan, G. A. and Marcq, J.},
  month   = may,
  year    = {2022},
  pages   = {20210167}
}

@article{liu_simulation_2021,
  title     = {Simulation of a generalized asset exchange model with economic growth and wealth distribution},
  volume    = {104},
  url       = {https://link.aps.org/doi/10.1103/PhysRevE.104.014150},
  doi       = {10.1103/PhysRevE.104.014150},
  number    = {1},
  journal   = {Phys. Rev. E},
  publisher = {American Physical Society},
  author    = {Liu, Kang K. L. and Lubbers, N. and Klein, W. and Tobochnik, J. and Boghosian, B. M. and Gould, Harvey},
  month     = jul,
  year      = {2021},
  pages     = {014150}
}

@article{newman_power_2005,
  title     = {Power laws, {Pareto} distributions and {Zipf}'s law},
  volume    = {46},
  issn      = {0010-7514},
  doi       = {10.1080/00107510500052444},
  number    = {5},
  journal   = {Contemporary Physics},
  publisher = {Taylor \& Francis},
  author    = {Newman, M. E. J.},
  month     = sep,
  year      = {2005},
  pages     = {323--351}
}

@article{peterson_nonuniversal_2010,
  title   = {Nonuniversal power law scaling in the probability distribution of scientific citations},
  volume  = {107},
  doi     = {doi:10.1073/pnas.1010757107},
  number  = {37},
  journal = {Proceedings of the National Academy of Sciences},
  author  = {Peterson, George J. and Pressé, Steve and Dill, Ken A.},
  year    = {2010},
  pages   = {16023--16027}
}

@article{merton_matthew_1968,
  title   = {The {Matthew} {Effect} in {Science}: {The} reward and communication systems of science are considered},
  volume  = {159},
  issn    = {0036-8075, 1095-9203},
  url     = {https://www.science.org/doi/10.1126/science.159.3810.56},
  doi     = {10.1126/science.159.3810.56},
  number  = {3810},
  journal = {Science},
  author  = {Merton, Robert K.},
  month   = jan,
  year    = {1968},
  pages   = {56--63}
}

@article{reed_pareto_2001,
  title   = {The {Pareto}, {Zipf} and other power laws},
  volume  = {74},
  issn    = {0165-1765},
  doi     = {https://doi.org/10.1016/S0165-1765(01)00524-9},
  number  = {1},
  journal = {Economics Letters},
  author  = {Reed, William J.},
  month   = dec,
  year    = {2001},
  pages   = {15--19}
}

@article{gabaix_zipfs_1999,
  title   = {Zipf's {Law} for {Cities}: {An} {Explanation}},
  volume  = {114},
  issn    = {0033-5533, 1531-4650},
  url     = {https://academic.oup.com/qje/article-lookup/doi/10.1162/003355399556133},
  doi     = {10.1162/003355399556133},
  number  = {3},
  journal = {The Quarterly Journal of Economics},
  author  = {Gabaix, X.},
  month   = aug,
  year    = {1999},
  pages   = {739--767}
}

@article{dragulescu_statistical_2000,
  title   = {Statistical mechanics of money},
  volume  = {17},
  issn    = {1434-6036},
  doi     = {10.1007/s100510070114},
  number  = {4},
  journal = {The European Physical Journal B - Condensed Matter and Complex Systems},
  author  = {Dragulescu, A. and Yakovenko, V. M.},
  month   = oct,
  year    = {2000},
  pages   = {723--729}
}

@article{bennati_il_1993,
  title   = {Il metodo di {Montecarlo} nell’analisi economica},
  volume  = {10},
  number  = {4},
  journal = {Rassegna di Lavori dell'ISCO (Istituto Nazionale per lo Studio della Congiuntura)},
  author  = {Bennati, E},
  year    = {1993},
  pages   = {31--79}
}

@article{sornette_dragon-kings_2012,
  title   = {Dragon-kings: {Mechanisms}, statistical methods and empirical evidence},
  volume  = {205},
  issn    = {1951-6355, 1951-6401},
  url     = {http://link.springer.com/10.1140/epjst/e2012-01559-5},
  doi     = {10.1140/epjst/e2012-01559-5},
  number  = {1},
  journal = {The European Physical Journal Special Topics},
  author  = {Sornette, D. and Ouillon, G.},
  month   = may,
  year    = {2012},
  pages   = {1--26}
}

@article{ritort_glassy_2003,
  title   = {Glassy dynamics of kinetically constrained models},
  volume  = {52},
  issn    = {0001-8732, 1460-6976},
  url     = {http://www.tandfonline.com/doi/abs/10.1080/0001873031000093582},
  doi     = {10.1080/0001873031000093582},
  number  = {4},
  journal = {Advances in Physics},
  author  = {Ritort, F. and Sollich, P.},
  month   = jun,
  year    = {2003},
  pages   = {219--342}
}

@article{bialas_condensation_1997,
  title   = {Condensation in the {Backgammon} model},
  volume  = {493},
  issn    = {05503213},
  url     = {https://linkinghub.elsevier.com/retrieve/pii/S0550321397001922},
  doi     = {10.1016/S0550-3213(97)00192-2},
  number  = {3},
  journal = {Nuclear Physics B},
  author  = {Bialas, Piotr and Burda, Zdzislaw and Johnston, Des},
  month   = jun,
  year    = {1997},
  pages   = {505--516}
}

@article{bouchaud_wealth_2000,
  title   = {Wealth condensation in a simple model of economy},
  volume  = {282},
  issn    = {03784371},
  url     = {https://linkinghub.elsevier.com/retrieve/pii/S0378437100002053},
  doi     = {10.1016/S0378-4371(00)00205-3},
  number  = {3-4},
  journal = {Physica A: Statistical Mechanics and its Applications},
  author  = {Bouchaud, Jean-Philippe and Mézard, Marc},
  month   = jul,
  year    = {2000},
  pages   = {536--545}
}

@article{burda_wealth_2002,
  title   = {Wealth condensation in {P}areto macroeconomies},
  volume  = {65},
  issn    = {1063-651X, 1095-3787},
  url     = {https://link.aps.org/doi/10.1103/PhysRevE.65.026102},
  doi     = {10.1103/PhysRevE.65.026102},
  number  = {2},
  journal = {Physical Review E},
  author  = {Burda, Z. and Johnston, D. and Jurkiewicz, J. and Kamiński, M. and Nowak, M. A. and Papp, G. and Zahed, I.},
  month   = jan,
  year    = {2002},
  pages   = {026102}
}

@article{boghosian_oligarchy_2017,
  title   = {Oligarchy as a phase transition: {The} effect of wealth-attained advantage in a {Fokker}–{Planck} description of asset exchange},
  volume  = {476},
  issn    = {03784371},
  url     = {https://linkinghub.elsevier.com/retrieve/pii/S037843711730081X},
  doi     = {10.1016/j.physa.2017.01.071},
  journal = {Physica A: Statistical Mechanics and its Applications},
  author  = {Boghosian, Bruce M. and Devitt-Lee, Adrian and Johnson, Merek and Li, Jie and Marcq, Jeremy A. and Wang, Hongyan},
  month   = jun,
  year    = {2017},
  pages   = {15--37}
}

@article{RosenResnick1980,
  author  = {Rosen, Kenneth T. and Resnick, Mitchel},
  title   = {The size distribution of cities: An examination of the {Pareto} law and primacy},
  journal = {Journal of Urban Economics},
  year    = {1980},
  volume  = {8},
  number  = {2},
  pages   = {165--186},
  doi     = {10.1016/0094-1190(80)90043-1}
}

@article{Soo2005,
  author  = {Soo, Kwok Tong},
  title   = {{Zipf}'s Law for cities: A cross-country investigation},
  journal = {Regional Science and Urban Economics},
  year    = {2005},
  volume  = {35},
  number  = {3},
  pages   = {239--263},
  doi     = {10.1016/j.regsciurbeco.2004.04.004}
}

@article{pisarenko_robust_2012,
  title   = {Robust statistical tests of {Dragon}-{Kings} beyond power law distributions},
  volume  = {205},
  issn    = {1951-6355, 1951-6401},
  url     = {http://link.springer.com/10.1140/epjst/e2012-01564-8},
  doi     = {10.1140/epjst/e2012-01564-8},
  number  = {1},
  journal = {The European Physical Journal Special Topics},
  author  = {Pisarenko, V. F. and Sornette, D.},
  month   = may,
  year    = {2012},
  pages   = {95--115}
}

@misc{UNWUP2025,
  author       = {{United Nations, Department of Economic and Social Affairs, Population Division}},
  title        = {{World Urbanization Prospects}: The 2025 Revision, Online Edition},
  year         = {2025},
  howpublished = {File 21: DEGURBA urban-centre populations},
  url          = {https://population.un.org/wup/downloads?tab=Cities}
}

@article{dragulescu_exponential_2001,
  title   = {Exponential and power-law probability distributions of wealth and income in the {United} {Kingdom} and the {United} {States}},
  volume  = {299},
  url     = {https://www.sciencedirect.com/science/article/pii/S0378437101002989},
  doi     = {https://doi.org/10.1016/S0378-4371(01)00298-9},
  number  = {1},
  journal = {Physica A},
  author  = {Drăgulescu, Adrian and Yakovenko, Victor M.},
  year    = {2001},
  pages   = {213--221}
}

@misc{cao_mean-field_2025,
  title         = {Mean-field analysis of a random asset exchange model with probabilistic cheaters},
  author        = {Cao, F.},
  year          = {2025},
  eprint        = {2505.06767},
  archivePrefix = {arXiv}
}

@article{blom_hallmarks_2025,
  title     = {Hallmarks of deception in asset-exchange models},
  volume    = {7},
  doi       = {10.1103/PhysRevResearch.7.013279},
  number    = {1},
  journal   = {Physical Review Research},
  publisher = {American Physical Society},
  author    = {Blom, Kristian and Makarov, Dmitrii E. and Godec, Aljaž},
  year      = {2025},
  pages     = {013279}
}

@article{boghosian_fokkerplanck_2014,
  title   = {Fokker–{Planck} description of wealth dynamics and the origin of {Pareto}'s law},
  volume  = {25},
  issn    = {0129-1831, 1793-6586},
  url     = {https://www.worldscientific.com/doi/abs/10.1142/S0129183114410083},
  doi     = {10.1142/S0129183114410083},
  number  = {12},
  journal = {International Journal of Modern Physics C},
  author  = {Boghosian, Bruce},
  month   = dec,
  year    = {2014},
  pages   = {1441008}
}

@article{sinha_stochastic_2003,
  title   = {Stochastic {Maps}, {Wealth} {Distribution} in {Random} {Asset} {Exchange} {Models} and the {Marginal} {Utility} of {Relative} {Wealth}},
  volume  = {T106},
  issn    = {0031-8949},
  url     = {https://iopscience.iop.org/article/10.1238/Physica.Topical.106a00059},
  doi     = {10.1238/Physica.Topical.106a00059},
  number  = {1},
  journal = {Physica Scripta},
  author  = {Sinha, Sitabhra},
  year    = {2003},
  pages   = {59}
}

@article{aydiner_universal_2023,
  title   = {Universal {Pareto} laws in agent-based exchange models: debt and varying initial-money distributions},
  volume  = {96},
  issn    = {1434-6028, 1434-6036},
  url     = {https://link.springer.com/10.1140/epjb/s10051-023-00579-y},
  doi     = {10.1140/epjb/s10051-023-00579-y},
  number  = {9},
  journal = {The European Physical Journal B},
  author  = {Aydiner, Ekrem and Cherstvy, Andrey G. and Metzler, Ralf and Sokolov, Igor M.},
  month   = sep,
  year    = {2023},
  pages   = {123}
}

@article{lucente_h-theorem_2025,
  title     = {H-theorem at negative temperature: the random exchange model with bounds},
  volume    = {2025},
  issn      = {1742-5468},
  doi       = {10.1088/1742-5468/ada49b},
  number    = {1},
  journal   = {Journal of Statistical Mechanics: Theory and Experiment},
  publisher = {IOP Publishing},
  author    = {Lucente, Dario and Baldovin, Marco and Puglisi, Andrea and Vulpiani, Angelo},
  month     = jan,
  year      = {2025},
  pages     = {013210}
}

@misc{lanchier_distribution_2022,
  title  = {Distribution of money on connected graphs with multiple banks},
  url    = {https://arxiv.org/abs/2201.11930},
  author = {Lanchier, Nicolas and Reed, Stephanie},
  year   = {2022}
}

@article{sinha_evidence_2006,
  title   = {Evidence for power-law tail of the wealth distribution in {India}},
  volume  = {359},
  issn    = {03784371},
  url     = {https://linkinghub.elsevier.com/retrieve/pii/S0378437105004826},
  doi     = {10.1016/j.physa.2005.02.092},
  journal = {Physica A: Statistical Mechanics and its Applications},
  author  = {Sinha, Sitabhra},
  month   = jan,
  year    = {2006},
  pages   = {555--562}
}

@article{coelho_double_2008,
  title   = {Double power laws in income and wealth distributions},
  volume  = {387},
  issn    = {03784371},
  url     = {https://linkinghub.elsevier.com/retrieve/pii/S0378437108000794},
  doi     = {10.1016/j.physa.2008.01.047},
  number  = {15},
  journal = {Physica A: Statistical Mechanics and its Applications},
  author  = {Coelho, Ricardo and Richmond, Peter and Barry, Joseph and Hutzler, Stefan},
  month   = jun,
  year    = {2008},
  pages   = {3847--3851}
}

@article{yukalov_statistical_2012,
  title   = {Statistical outliers and dragon-kings as {Bose}-condensed droplets},
  volume  = {205},
  issn    = {1951-6355, 1951-6401},
  url     = {http://link.springer.com/10.1140/epjst/e2012-01561-y},
  doi     = {10.1140/epjst/e2012-01561-y},
  number  = {1},
  journal = {The European Physical Journal Special Topics},
  author  = {Yukalov, V. I. and Sornette, D.},
  month   = may,
  year    = {2012},
  pages   = {53--64}
}

@article{KaplanMeier1958,
  author  = {Kaplan, E. L. and Meier, Paul},
  title   = {Nonparametric Estimation from Incomplete Observations},
  journal = {Journal of the American Statistical Association},
  year    = {1958},
  volume  = {53},
  number  = {282},
  pages   = {457--481},
  doi     = {10.1080/01621459.1958.10501452}
}

@article{Wilson1927,
  author  = {Wilson, Edwin B.},
  title   = {Probable Inference, the Law of Succession, and Statistical Inference},
  journal = {Journal of the American Statistical Association},
  year    = {1927},
  volume  = {22},
  number  = {158},
  pages   = {209--212},
  doi     = {10.1080/01621459.1927.10502953}
}

@article{BrownCaiDasGupta2001,
  author  = {Brown, Lawrence D. and Cai, T. Tony and DasGupta, Anirban},
  title   = {Interval Estimation for a Binomial Proportion},
  journal = {Statistical Science},
  year    = {2001},
  volume  = {16},
  number  = {2},
  pages   = {101--133},
  doi     = {10.1214/ss/1009213286}
}

\clearpage
\onecolumngrid

\setcounter{section}{0}
\setcounter{equation}{0}
\setcounter{figure}{0}
\setcounter{table}{0}
\renewcommand{\thesection}{S\arabic{section}}
\renewcommand{\theequation}{S\arabic{equation}}
\renewcommand{\thefigure}{S\arabic{figure}}
\renewcommand{\thetable}{S\arabic{table}}
\renewcommand{\theHsection}{S\arabic{section}}
\renewcommand{\theHequation}{S\arabic{equation}}
\renewcommand{\theHfigure}{S\arabic{figure}}
\renewcommand{\theHtable}{S\arabic{table}}

\begin{center}
{\large\bfseries Supporting Information}
\end{center}
\vspace{1ex}

\section{Continuum solution for poor-biased exchange}
\label{SI-sec:ContSol}
Here we consider the case $-\frac{1}{2}<\eta_\infty <0$, a case where transfer from the more wealthy to the less wealthy transaction partner is favored. Using Eq.~\ref{key approx}, we rewrite Eq.~\ref{interation general} as
\begin{equation} \label{SI-iteration poor-biased}
    \frac{p(m+1)}{p(m)}\approx\frac{\frac{1}{2}+2\eta_\infty [\frac{1}{2}-S(m)]-p(0)(\frac{1}{2}+\eta_\infty)}{\frac{1}{2}-2\eta_\infty [\frac{1}{2}-S(m)]}.
\end{equation}
If $|\eta_\infty|\ll p(0)$, the wealth distribution
approaches the fair-game result,
$p(m+1)/p(m)\approx 1-p(0)$. We therefore focus on the regime in which
$p(0)$ and $|\eta_\infty|$ are comparable,
or $p(0)\ll|\eta_\infty|$.
We additionally assume $|\eta_\infty|\ll1$,
corresponding to weak exchange bias. In this case, we can  approximate Eq.~\ref{SI-iteration poor-biased} with
\begin{equation}
     \frac{p(m+1)}{p(m)}\approx 1+8\eta_\infty [\frac{1}{2}-S(m)]-p(0),
\end{equation}
This equation can be approximated, in the continuum limit,  as (cf.  Eq.~\ref{diff E for distribution})
\begin{equation}  \label{SI-logarithmic decrease}
    \frac{d \ln p(m)}{dm} = -\beta [S(m)],
\end{equation}
where
\begin{equation} \label{SI-beta for poor}
    \beta(S)=8|\eta_\infty|(\frac{1}{2}-S)+p(0).
\end{equation}
This function characterizes how the rate of logarithmic wealth decline depends on the tail probability. Because this type of equation is encountered here more than once (for example, setting $\beta(S)$ to a constant results in an exponential distribution, $p(m)\propto e^{-\beta m}$, while the case $\beta(S)\propto S$ results in a power law $p(m)\propto m^{-2}$ described in the main text (Eq.~\ref{Pareto})), it is worth writing down its general solution.  For an arbitrary function $\beta(S)$ we can integrate this equation as follows: Using Eq.~\ref{relation between S and p},  Eq.~\ref{SI-logarithmic decrease} can be written as
\begin{equation} \label{SI-differential eq for S}
    \frac{d^2S}{dm^2}=-\frac{d \gamma[S(m)]}{dm},
\end{equation}
where
\begin{equation} \label{SI-gamma}
    \gamma(S)=\int^S \beta(S') dS'
\end{equation}
Choosing the  integration constant such that $\gamma(0)=0$, we obtain the relationship between $S$ and $m$ (i.e., solution of Eq.~\ref{SI-differential eq for S}) in the form
\begin{equation} \label{SI-integrate gamma}
    \int^S \frac{dS'}{\gamma(S')}=-(m-m_0
)
\end{equation}
with an appropriately chosen integration constant $m_0$. For $\beta(S)$ from Eq.~\ref{SI-beta for poor},
using Eqs.~\ref{SI-gamma} and \ref{SI-integrate gamma},  and choosing the integration constant $m_0$ such that $S(0)=1$, one obtains
\begin{equation}
    S(m)=\frac{4|\eta_\infty|+p(0)}{p(0) e^{[4|\eta_\infty|+p(0)]m}+4|\eta_\infty|}
\end{equation}
and
\begin{equation} \label{SI-theory poor biased}
    p(m)=-\frac{dS(m)}{dm} = \frac{[4|\eta_\infty|+p(0)]^2p(0)e^{[4|\eta_\infty|+p(0)]m}}{\{p(0) e^{[4|\eta_\infty|+p(0)]m}+4|\eta_\infty|\}^2}.
\end{equation}
Notice that, although one does not necessarily expect that the continuous approximation would work for small values of $m$, this result passes the consistency test, recovering $p(0)$ when $m=0$.  Finally, we estimate the expected value of wealth per player,
\begin{equation}
    \langle m \rangle = \int_0^\infty m p(m) dm = \frac{1}{4|\eta_\infty|} \ln[1+\frac{4|\eta_\infty|}{p(0)}],
\end{equation}
from which we express $p(0)$ as
\begin{equation}\label{SI-mean wealth poor biased theory}
    p(0)=\frac{4|\eta_\infty|}{e^{4|\eta_\infty|\langle m\rangle }-1}.
\end{equation}
In particular, assuming $p(0)\ll |\eta_\infty|$, we obtain an especially simple approximation for the distribution   $p(m)$:
\begin{equation} \label{SI-especially simple}
    p(m)\approx \frac{|\eta_\infty|}{\cosh^2[2 |\eta_\infty|(m-\langle m\rangle)]},
\end{equation}
a function that is peaked at the mean value $\langle m\rangle$, with an exponential tail.  This is illustrated in Fig.~\ref{fig:poor-biased-data}, which shows wealth distributions obtained using our analytical theory and with Monte Carlo simulations of the game with a large number of players, showing excellent agreement.

\section{Justification of Eq.~\ref{key approx}}
\label{SI-sec:asymp_just}
We write Eq.~(8) as
\begin{equation}
f(m)
=
\sum_{m'=0}^{m-1} p(m')\eta(m'-m)
+
\sum_{m'=m+1}^{\infty} p(m')\eta(m'-m),
\label{SI-A1}
\end{equation}
where the term with $m'=m$ vanishes because $\eta(0)=0$. For positive integers $j$, define
\begin{equation}
\Delta(j)\equiv \eta_\infty-\eta(j).
\label{SI-A2}
\end{equation}
Since $\eta$ is odd,
\[
\eta(-j)=-\eta_\infty+\Delta(j).
\]
We assume that, for sufficiently large $j$,
\begin{equation}
|\Delta(j)|\leq \frac{C_\Delta}{j^k},
\qquad k>1,
\label{SI-A3}
\end{equation}
where $C_\Delta$ is a constant.

Substituting Eq.~(\ref{SI-A2}) into Eq.~(\ref{SI-A1}) gives
\begin{equation}
f(m)
=
\eta_\infty[2S(m)-1-p(m)]
+
E_-(m)-E_+(m),
\label{SI-A4}
\end{equation}
where
\begin{equation}
\begin{split}
E_-(m)
&= \sum_{m'=0}^{m-1}p(m')\Delta(m-m'),\\
E_+(m)
&=
\sum_{m'=m+1}^{\infty}p(m')\Delta(m'-m).
\label{SI-A5}
\end{split}
\end{equation}
Here we used $S(m+1)=S(m)-p(m)$.

We now estimate the two error terms, $E_{\pm}(m)$, using the critical asymptotic behavior
\[
p(m)=O(m^{-2}),
\qquad
S(m)=O(m^{-1}).
\]
First, split $E_-(m)$ at $m/2$:
\begin{equation}
\begin{split}
E_-(m)
&=
\sum_{m'=0}^{\lfloor m/2\rfloor}
p(m')\Delta(m-m')
+
\sum_{m'=\lfloor m/2\rfloor+1}^{m-1}
p(m')\Delta(m-m')
\equiv E_-^{(1)}+E_-^{(2)}.
\label{SI-A6}
\end{split}
\end{equation}
For the first term, $m-m'\geq m/2$, and therefore
\begin{equation}
|E_-^{(1)}|
\leq
C_\Delta\left(\frac{m}{2}\right)^{-k}
\sum_{m'=0}^{\lfloor m/2\rfloor}p(m')
\leq
C_\Delta\,2^k m^{-k},
\label{SI-A7}
\end{equation}
so that $E_-^{(1)}=O(m^{-k})$.

For the second term, set $j=m-m'$. Since $m-j\geq m/2$,
\begin{equation}
|E_-^{(2)}|
\leq
\frac{C_p}{m^2}
\sum_{j=1}^{\lfloor m/2\rfloor}|\Delta(j)|
=
O(m^{-2}),
\label{SI-A8}
\end{equation}
where $C_p$ is a constant. The last step follows because Eq.~(\ref{SI-A3}) with $k>1$ implies that $\sum_j|\Delta(j)|$ converges.

Similarly, changing variables to $j=m'-m$ in $E_+(m)$ gives
\begin{equation}
|E_+(m)|
\leq
\frac{C_p}{m^2}
\sum_{j=1}^{\infty}|\Delta(j)|
=
O(m^{-2}).
\label{SI-A9}
\end{equation}

Combining these estimates with Eq.~(\ref{SI-A4}), and using $p(m)=O(m^{-2})$, we obtain
\begin{equation}
f(m)
=
\eta_\infty[2S(m)-1]
+
O(m^{-k})
+
O(m^{-2}).
\label{SI-A10}
\end{equation}
For $k>1$, both corrections are $o(m^{-1})$, whereas $S(m)=O(m^{-1})$. Hence
\begin{equation}
f(m)
=
\eta_\infty[2S(m)-1]
+
o(m^{-1}),
\qquad m\rightarrow\infty.
\label{SI-A11}
\end{equation}

Thus Eq.~(\ref{key approx}) is asymptotically valid to the order required to determine the distribution $p(m)$, at large $m$ and in the critical case, whenever $\eta(n)$ approaches its limiting values faster than $1/n$. In particular, for
\[
|\eta_\infty-\eta(n)|=O(n^{-k}),
\qquad k>1,
\]
the detailed form of the interaction rule contributes only subleading corrections, making the $p(m)\propto m^{-2}$ solution of Eq.~(\ref{relation between S and p}) asymptotically self-consistent.

\section{Near-critical distribution and exponential crossover}
\label{SI-sec:NearCrit}
Here we estimate the wealth distribution in the vicinity of the critical point, where $p(0)$ approaches the critical value $p_c$ from above. In this case, Eq.~\ref{interation general} becomes
\begin{equation}
    \frac{p({m+1})}{p(m)} \approx \frac{1-p(0)-p_c S(m)}{1-p_c +p_c S(m)}\approx 1-\delta-\frac{4\eta_\infty}{\frac{1}{2}-\eta_\infty} S(m),
\end{equation}
where
\begin{equation} \label{SI-delta}
\delta =\frac{p(0)-p_c}{1-p_c}.
\end{equation}

In the continuum limit, this becomes Eq.~\ref{SI-logarithmic decrease} with
\[
\beta(S) = \delta + \frac{4\eta_\infty}{\frac{1}{2}-\eta_\infty} S.
\]
Using Eqs.~\ref{SI-gamma} and \ref{SI-integrate gamma}, one finds
\begin{equation} \label{SI-tail subcritical}
S(m)=\frac{\kappa\delta}{e^{\delta(m-m_0)}-1},
\end{equation}
where $m_0$ is an integration constant. Differentiating Eq.~\ref{SI-tail subcritical} gives
\begin{equation} \label{SI-distribution subcritical}
    p(m)=-S'(m)\approx \frac{\delta^2 \kappa}{4 \sinh^2[\delta (m-m_0)/2]}.
\end{equation}
The integration constant $m_0$ is a matching offset that may depend on
$\delta$. In the range $|m_0|\ll m\ll\delta^{-1}$, the continuum density
reduces to $p(m)\simeq\kappa/(m-m_0)^2\sim\kappa/m^2$. At larger wealth,
where $\delta(m-m_0)\gg1$, the density decays exponentially as
$e^{-\delta m}$.

For Fig.~3B, we fitted $m_0$ separately for each value of $\delta$,
keeping $\delta$ and $\kappa$ fixed. We minimized the mean squared
difference between the log densities of the continuum expression and
the numerical mean-field solution over the fitting range specified in
the SI Appendix.

\section{Finite-population mean-field theory with a separated outlier}
\label{SI-sec:finite_bulk}
Consider a finite population containing one well-separated wealthy outlier and $N-1$ remaining agents, with $0<\eta_\infty<1/2$. Using analogy with physical condensation phenomena, we can think of those $N-1$ agents as forming the ``bulk'' of the system, terminology that will be used below. We consider the regime in which the outlier is sufficiently wealthy that its interactions with the remaining agents have reached their asymptotic exchange probabilities, and we neglect correlations among the remaining agents.
Throughout this appendix, $p(m)$ and $p(m'|m)$ refer to a bulk
agent: $\sum_{m=0}^\infty p(m)=1$, and $f(m)$ is evaluated from this
distribution using Eq.~\ref{f function}. The condition that the outlier is in steady-state balance with the rest of the agents,  Eq.~\ref{critical Z},  gives
$p(0)=p_c$ within this approximation.  The fraction of the agents with zero wealth within the entire
population including the outlier is $(1-1/N)p_c$.
A bulk agent encounters another bulk agent with probability
$(N-2)/(N-1)$ and the outlier with probability $1/(N-1)$.
Its gain and loss probabilities in such an encounter are, therefore,
\begin{equation}
\begin{aligned}
p(m+1| m)&=\frac{N-2}{N-1}
\left[\frac{1-p_c}{2}-f(m)-p_c\eta(m)\right]
+\frac{\frac12-\eta_\infty}{N-1},\qquad m\geq0,\\
p(m-1| m)&=\frac{N-2}{N-1}\left[\frac12+f(m)\right]
+\frac{\frac12+\eta_\infty}{N-1},\qquad m\geq1,
\end{aligned}
\label{SI-eq:finite_bulk_rates}
\end{equation}
with $p(-1|0)=0$. The subtraction of $p_c\eta(m)$, together with
the factor $1-p_c$, excludes empty bulk donors from the gain probability.

The steady-state condition is Eq.~\ref{DB},
$p(m+1)/p(m)=p(m+1|m)/p(m| m+1)$.
At fixed $N$, we first assume that $M$ is large enough that the wealthy outlier is well separated from the rest of the population. We then examine the large-$m$ asymptotics of the resulting bulk distribution, where we have
$f(m)\to-\eta_\infty$ and
$\eta(m)\to\eta_\infty$. This results in
\begin{equation}
\frac{p(m+1)}{p(m)}\longrightarrow
\frac{1}{1+1/[\kappa(N-1)]}<1.
\label{SI-eq:finite_bulk_tail_ratio}
\end{equation}
Here, $\kappa$ is defined in the main text, Eq.~\ref{interation critical}.
Thus, the bulk has an exponential tail with the decay length
$\{\ln[1+1/(\kappa(N-1))]\}^{-1}\simeq\kappa(N-1)$,
where the approximation holds for $\kappa(N-1)\gg1$.
At large $N$, a critical Pareto regime can precede this cutoff.
The decay length is approximately equal to $m^*$, consistent with
the outlier's influence becoming important where the critical tail
contains approximately one agent.

The exponential cutoff keeps the bulk mean finite even as
$M\to\infty$ at fixed $N$. Once the
outlier interactions saturate, the bulk equations contain no further
dependence on $M$. Their stationary solution therefore predicts
\begin{equation}
\frac{M-M_{DK}}{N}
\simeq\frac{N-1}{N}\sum_{m=0}^\infty m\,p(m),
\label{SI-eq:finite_bulk_partition}
\end{equation}
where \[M_{\rm DK}\equiv \langle \max_i (m_i) \rangle\] is the stationary average of the
largest wealth, as in the main text. The right-hand side is independent of
further increases in $M$ at fixed $N$.
This gives an approximately constant bulk contribution and
near-unit-slope growth of $M_{DK}/N$ against
$\langle m\rangle=M/N$, as tested in Fig.~\ref{fig:finiteN-one-outlier}F.

\section{Additional numerical methods and convergence tests}
\label{SI-sec:Num}

\subsection*{Exchange dynamics and simulation protocols}
All Monte Carlo simulations conserve the total wealth. In each sweep, agents are randomly permuted and grouped into \(N/2\) disjoint interacting pairs. Exchange probabilities are determined by the bias functions defined in the main text, with \(\eta(0)=0\).
Model parameters and numerical schedules
are summarized in Tables~\ref{SI-tab:SI_parameters}--\ref{SI-tab:SI_EF};
parameters essential for interpreting individual results are also given in the corresponding
figure captions.

Figure~2 uses equal-wealth initial conditions. Figure~3A
uses stratified sampling from the stationary mean-field
distribution, with the sampled wealths adjusted to conserve
the prescribed total wealth.

Figure~\ref{fig:finiteN-one-outlier}A,B follows
the emergence of condensation from equal-wealth initial
conditions, with individual trajectories identified by
their final wealth ranks. Panel C examines the relaxation
of a precondensed state prepared by adding excess wealth
to one agent in an equilibrated population. The background
is equilibrated before enrichment, and the enriched
population is equilibrated again before measurements.
Panel D uses independent stationary trajectories at each
mean wealth.

\subsection*{Observables and uncertainty}
The stationary one-agent distribution $p_N(m)$ is the time average of
snapshot histograms normalized by $N$; its survival function is
$S_N(m)=\sum_{u=m}^{M}p_N(u)$.
Figure~\ref{fig:finiteN-one-outlier}D plots $mS_N(m)/\kappa$.
Removing the largest wealth in each snapshot defines the normalized bulk
distribution $B(m)$ and the removed-maximum distribution $D(m)$, so
\begin{equation}
p_N(m)=\frac{N-1}{N}B(m)+\frac1N D(m).
\end{equation}
The identity of the maximum can change over time.
Figure~\ref{fig:finiteN-one-outlier}F displays
$M_{DK}/N$ and $(M-M_{DK})/N$;
these sum to $\langle m\rangle$ and are not conditional means per bulk agent.
As the mean wealth increases from 6 to 12, the richest
agent accumulates $99.6\pm0.8\%$ of the additional wealth
(mean $\pm$ standard error).

Error bars are standard errors across independent
trajectory averages, rather than correlated observations
within individual trajectories.

\subsection*{Stationary mean-field calculations}
For a trial distribution $p(m)$, the interaction term
$f(m)$ is evaluated using Eq.~\ref{f function}.
The one-unit gain and loss probabilities are
\begin{align}
p(m+1| m)
&=\frac{1-p(0)}{2}-f(m)-p(0)\eta(m),
\qquad m\geq0,\\
p(m-1| m)
&=\frac12+f(m),
\qquad m\geq1,
\end{align}
with $p(-1|0)=0$.
Zero adjacent-state flux gives
\begin{equation}
p(m+1)=p(m)
\frac{p(m+1 | m)}{p(m | m+1)},
\end{equation}
consistent with Eq.~\ref{DB}.
The distribution and interaction term are iterated
self-consistently on a finite wealth domain, with
normalization at each step. Convergence is assessed
from successive changes, flux residuals including
the zero-wealth boundary, and domain comparisons
(Table~\ref{SI-tab:SI_MF}).
For Figs.~3A--B, an outer bisection adjusts $p(0)$
to obtain the prescribed mean wealth. For Fig.~3C,
$p(0)$ is specified through
$\delta=[p(0)-p_c]/(1-p_c)$.

\subsection*{Comparisons with continuum approximations and finite-size crossover}
Figure~3B compares numerical mean-field distributions
with the near-critical continuum approximation
(Eq.~\ref{SI-distribution subcritical}). Only the matching
offset $m_0$ is fitted, by minimizing squared differences
between the logarithms of the two distributions over
$20\leq m\leq200$ with $p_{\mathrm{MF}}(m)\geq10^{-8}$;
$\delta$ and $\kappa$ remain fixed.
Figure~2 uses Eq.~\ref{SI-theory poor biased}, with $p(0)$ fixed by the mean wealth through Eq.~\ref{SI-mean wealth poor biased theory}; no parameters are fitted.

The crossover between the uncondensed and condensed
regimes is estimated from the finite-population capacity
of the critical bulk distribution, from Eq.~\ref{eq:crossover_matching}, i.e. \( m_{mx} =\kappa\ln(\kappa N).\)
For the step bias with $\eta_\infty=1/4$,
$\kappa=1/2$, and thus
$m_{mx}\approx2.505$ for $N=300$.
This estimate is shown in
Fig.~\ref{fig:finiteN-one-outlier}E--F.
The finite-population mean-field theory in
SI Section~\ref{SI-sec:finite_bulk} describes the
stationary bulk distribution in the condensed regime.

\subsection*{Near-critical convergence and reproducibility}

For Fig.~3C, $\delta=10^{-5}$ places the subcritical cutoff
beyond the plotted range. At $m=20{,}000$, the leading
continuum cutoff factor differs from the critical envelope
by approximately $0.33\%$. Each of the four bias functions
was solved independently on domains $L=800{,}000$ and
$1{,}600{,}000$. Over $1\leq m\leq20{,}000$, the largest
relative difference between the two solutions, using the
larger-domain solution as the denominator, was
$4.51\times10^{-6}$ across the four bias functions.
Iteration counts and stationary-flux residuals are reported
in Table~\ref{SI-tab:SI_convergence}. These comparisons
demonstrate numerical convergence over the plotted range.

The accompanying reproducibility package contains Python
code and parameter configurations for independently
recomputing the Monte Carlo simulations, stationary
mean-field solutions, analytical predictions, and figures.
The code uses NumPy, SciPy, and pandas for numerical calculations and
data processing, Numba for accelerated Monte Carlo updates,
and Matplotlib for plotting. Simulation parameters,
initialization procedures, sampling schedules, and random
seeds are specified in the code and accompanying tables.
The calculations generate numerical results and diagnostic
records, including checks of wealth conservation,
probability normalization, and numerical convergence.
Figures can therefore be regenerated from the model
equations and simulation procedures rather than from
previously saved datasets.
The resulting figures were subsequently edited in
Adobe Illustrator to improve clarity and presentation,
without altering the underlying numerical data or results.

\section{City-size statistical methods and robustness analyses}
\label{SI-sec:CityMethods}

\paragraph*{Data and selection}
We used 2025 population estimates from File~21 of the United Nations
\emph{World Urbanization Prospects 2025}, based on the harmonized Degree
of Urbanization (DEGURBA) definition of urban centers, thereafter cities~\cite{UNWUP2025}.
These spatial units differ from administrative municipalities and
commuting-based metropolitan areas. We retained cities with at least
50,000 inhabitants and countries with at least $K=40$ such cities,
yielding $C=53$ countries. Within country $c$ the largest 40 populations
are ordered $m_{c,1}\geq\cdots\geq m_{c,40}$ and normalized by
$x_{c,j}=m_{c,j}/m_{c,40}$. The largest city is excluded from the Pareto
``background'', defined here by ranks 2--40.

\paragraph*{Aggregate rank profile and pooled density}
Figure~5B uses the geometric mean at each rank,
\begin{equation}
g_k=\exp\!\left[\frac1C\sum_{c=1}^{C}\ln x_{c,k}\right].
\label{SI-eq:city_aggregate_rank}
\end{equation}
A log--log least-squares fit over ranks 2--40 gives
$g_k=32.02\,k^{-0.938}$; the observed rank-1 value is $g_1=59.83$.
Shading spans the 25th--75th percentiles across countries.
An independent Pareto-sampling simulation of this same aggregate
rank-1 statistic yielded values at least as large as observed in
82.21\% of $10^6$ repetitions. Thus aggregate rank-1 elevation alone
does not establish anomalous leading-city dominance.

For Fig. 5A, after withholding each country’s maximum, the second-largest city provides a right-censoring threshold: there are 38 observed exceedances and one right-censored observation. The pooled Kaplan–Meier (product-limit) estimate~\cite{KaplanMeier1958} was then converted to a density using 15 logarithmic bins.
For adjacent ranks $j=1,\ldots,39$, define weighted log spacings and their mean in the background,
\begin{equation}
W_{c,j}=j\ln\!\left(\frac{m_{c,j}}{m_{c,j+1}}\right),\qquad
\bar W_c=\frac{1}{38}\sum_{j=2}^{39}W_{c,j}.
\label{SI-eq:city_spacings}
\end{equation}
For a Pareto survival law \(P(X\ge x)\propto x^{-q}\), with survival exponent $q$, the background likelihood is
proportional to $(q\exp(-q\bar W_c))^{38}$, giving
\begin{equation}
\hat q=\frac{C}{\sum_c \bar W_c},\qquad
p(x)\propto x^{-(1+\hat q)}.
\label{SI-eq:city_pooled_exponent}
\end{equation}
For all 53 countries, $\hat q=1.09770$ (density exponent 2.09770).
A likelihood-ratio test of a common background exponent across countries
gives $p=2.51\times10^{-4}$ under the specified Pareto null. The null distribution of the likelihood-ratio statistic was calibrated by Monte Carlo sampling of the Gamma-distributed sums of the spacings expected from sums of independent exponential spacings in the background.

\paragraph*{Leading-city score and reference distribution}
Under independent Pareto sampling within each country, the weighted
log spacings are independent exponential variables with common rate
$q_c$, conditional on the 40th-ranked population. Following the
one-outlier dragon-king test~\cite{pisarenko_robust_2012}, we define
\begin{equation}
R_c=\frac{W_{c,1}}{\bar W_c},\qquad
R_c\sim F_{2,76},\qquad
p_{\mathrm{DK},c}=\left(1+\frac{R_c}{38}\right)^{-38}.
\label{SI-eq:city_DK}
\end{equation}
The unknown $q_c$ cancels, so the reference does not require a common
exponent. The null median is
$r_{1/2}=38(2^{1/38}-1)\simeq0.6995$.
The corresponding reference largest-city population is
\begin{equation}
m_{c,1}^{\mathrm{ref}}=m_{c,2}\exp\!\left(
 r_{1/2}{\bar W_c}\right).
\label{SI-eq:city_median_population}
\end{equation}
Thus $m_{c,1}>m_{c,1}^{\mathrm{ref}}$ is equivalent to
$R_c>r_{1/2}$ and has null probability $1/2$.
The resulting ratios of observed to reference
largest-city populations are 4.17 for Paris,
17.22 for Bangkok, 8.04 for Addis Ababa,
and 0.959 for S\~ao Paulo.

\paragraph*{Background power-law test}
We assess whether the smaller cities in each country are
consistent with a Pareto distribution, using only the
background spacings $W_{c,2},\ldots,W_{c,39}$.
Under the Pareto model, the cumulative normalized spacings
\[
U_{c,\ell}=
\frac{\sum_{j=2}^{\ell+1}W_{c,j}}{38\bar W_c},
\qquad \ell=1,\ldots,37,
\]
have the same distribution as 37 ordered independent
Uniform$(0,1)$ observations. We therefore apply the
two-sided Kolmogorov--Smirnov (KS) test,
\begin{equation}
D_c=\max_{1\leq\ell\leq37}
\left\{
\frac{\ell}{37}-U_{c,\ell},\quad
U_{c,\ell}-\frac{\ell-1}{37}
\right\},
\qquad
p_{\mathrm{fit},c}=1-F_{\mathrm{KS},37}(D_c),
\label{SI-eq:city_background_KS}
\end{equation}
where $F_{\mathrm{KS},37}$ is the cumulative distribution of
the KS statistic for 37 observations.
Requiring $p_{\mathrm{fit},c}\geq0.10$
retains 45 of the 53 countries.

Under the Pareto model, the normalized background
spacings are independent of the largest-city statistic
$R_c$. Thus, selecting countries based on this test
does not change the expected probability of $1/2$
that their largest-city statistic exceeds the
theoretical median.

\paragraph*{Statistical significance and robustness}
Under the Pareto model, each country's largest-city
statistic $R_c$ has a probability of $1/2$ of exceeding
the theoretical median $r_{1/2}$. Assuming independence
between countries, the number $A$ of countries exceeding
this threshold follows a binomial distribution, giving
the one-sided probability
\begin{equation}
p_{\mathrm{aggregate}}=
2^{-C}\sum_{a=A}^{C}\binom{C}{a},
\label{SI-eq:city_binomial}
\end{equation}
where $C$ is the number of countries.
We find that 36 out of 53 countries exceed  this threshold, resulting in $p=0.006330$;
or, after excluding those that fail
the background power-law test, 32 of 45 countries do so ($p=0.003304$).
Figure~5D shows the corresponding fractions with
95\% Wilson confidence intervals
\cite{Wilson1927,BrownCaiDasGupta2001}.
These results indicate an excess of dominant
largest cities relative to the Pareto sampling model.

Reducing the minimum number of cities per country
from 40 to 30 or 20 increases the sample from 53
to 61 or 79 countries. The pooled Pareto density
exponents remain close to 2 (2.098, 2.113, and
2.152, respectively). The numbers of countries
with largest-city scores above the theoretical
median are 36/53, 42/61, and 60/79, with
one-sided binomial $p$-values of 0.00633,
0.00222, and $2.10\times10^{-6}$.

\clearpage

\begin{table}[!htbp]
\centering\small
\setlength{\tabcolsep}{4pt}\renewcommand{\arraystretch}{1.15}
\caption{Model parameters for the numerical figures. $N$ is the number of
agents, not the number of trajectories. MF denotes an infinite-population
mean-field calculation.}
\label{SI-tab:SI_parameters}
\begin{tabular}{@{}p{0.09\linewidth} p{0.09\linewidth} p{0.36\linewidth} p{0.07\linewidth} p{0.32\linewidth}@{}}
\toprule
Panel & $N$ & Interaction & $\eta_\infty$ & Mean wealth or control \\
\midrule
2 & $10{,}000$ & Step, i.e. $\eta_\infty\operatorname{sgn}(n)$ & $-0.15$ & $\langle m\rangle=4,\ 10$ \\
3A & $5{,}000$ & Tanh, i.e. $\eta_\infty\tanh(n/3)$,  & $1/4$ & 6 means; Table~\ref{SI-tab:SI_MC} \\
3B & MF & Tanh,  & $1/4$ & $\langle m\rangle=1.4,\ 2.6,\ 3.1$\\
3C & MF & \raggedright Step; tanh; $\eta_\infty\operatorname{sgn}(n)|n|^h/(3^h+|n|^h),\  h=4,2$ & $1/4$ & $\delta=10^{-5}$;  for smooth rules \\
4A--B & $300$ & Step & $1/4$ &
$\langle m\rangle=5$; $M=1{,}500$ \\
4C & $300$ & Step & $1/4$ & $\langle m\rangle=4.03$; $M=1{,}209$ \\
4D & $300$ & Step & $1/4$ & $\langle m\rangle=1.38$, $2.12$, $2.97$, $4.03$ \\
4E--F & $300$ & Step & $1/4$ &
13 means; Table~\ref{SI-tab:SI_EF} \\
\bottomrule
\end{tabular}
\par\raggedright\footnotesize
\end{table}

\begin{table}[!htbp]
\centering\small
\setlength{\tabcolsep}{4pt}\renewcommand{\arraystretch}{1.12}
\caption{Monte Carlo simulation schedules outside the
Fig.~4E--F ensembles. $R$ is the number of independent
trajectories; $B$ is the equilibration duration, $T$ is
the measurement duration, and $\Delta t_{\rm samp}$ is the interval between
recorded configurations or observables, all measured in
sweeps. The integer total wealth $M$ specifies the exact
mean wealth $M/N$.}
\label{SI-tab:SI_MC}
\begin{tabular}{@{}l r r r r r r r@{}}
\toprule
Panel & $\langle m\rangle$ & $N$ & $M$ & $R$ & $B$ & $T$ & $\Delta t_{\rm samp}$ \\
\midrule
2 & 4 & 10,000 & 40,000 & 2 & 5,000 & 10,000 & 10 \\
2 & 10 & 10,000 & 100,000 & 2 & 5,000 & 10,000 & 10 \\
3A & 1.4224 & 5,000 & 7,112 & 12 & 8,000 & 16,000 & 100 \\
3A & 1.6230 & 5,000 & 8,115 & 12 & 8,000 & 16,000 & 100 \\
3A & 1.9688 & 5,000 & 9,844 & 12 & 10,000 & 20,000 & 100 \\
3A & 2.3158 & 5,000 & 11,579 & 12 & 12,000 & 24,000 & 100 \\
3A & 2.5716 & 5,000 & 12,858 & 12 & 16,000 & 32,000 & 100 \\
3A & 3.1218 & 5,000 & 15,609 & 12 & 24,000 & 48,000 & 100 \\
4A--B$^{a}$ & 5 & 300 & 1,500 & 1 & 0 & 3,000,000 & 1,000 \\
4C$^{b}$ & 4.03 & 300 & 1,209 & 4 & 300,000 & 1,200,000 & 500 \\
4D & 1.38 & 300 & 414 & 4 & 600,000 & 600,000 & 500 \\
4D & 2.12 & 300 & 636 & 4 & 600,000 & 600,000 & 500 \\
4D & 2.97 & 300 & 891 & 4 & 1,500,000 & 600,000 & 500 \\
4D & 4.03 & 300 & 1,209 & 4 & 1,500,000 & 600,000 & 500 \\
\bottomrule
\end{tabular}
\par\raggedright\footnotesize
$^{a}$For Figs.~4A--B, $T$ is the total simulation
duration. The empty fraction is recorded every
$1{,}000$ sweeps, with additional observations
at early logarithmically spaced times.
$^{b}$For Fig.~4C, a population with $M=636$ is
equilibrated for $10^6$ sweeps before adding
573 wealth units to an initially empty agent.
The tabulated schedule applies after enrichment.
\end{table}

\begin{table}[!htbp]
\centering\small
\setlength{\tabcolsep}{5pt}\renewcommand{\arraystretch}{1.12}
\caption{Fig.~4E,F production ensembles.
Every row uses $N=300$, $R=12$,
and snapshot interval $\Delta t_{\rm samp}=500$ sweeps. Mean wealth is shown to four decimal
places where necessary; $M$ specifies its exact value. E denotes a broad geometric initial wealth distribution;
P denotes an initial state in which excess wealth is
concentrated on a single agent. Both are adjusted to
the prescribed total wealth before equilibration.}
\label{SI-tab:SI_EF}
\begin{tabular}{@{}r r r r l@{}}
\toprule
$\langle m\rangle$ & $M$ & Equilibration $B$ & Production $T$ & Initial states \\
\midrule
1.38 & 414 & 800,000 & 800,000 & 12 E \\
1.59 & 477 & 800,000 & 800,000 & 12 E \\
1.8033 & 541 & 800,000 & 800,000 & 12 E \\
1.9733 & 592 & 800,000 & 800,000 & 12 E \\
2.12 & 636 & 800,000 & 800,000 & 12 E \\
2.25 & 675 & 1,200,000 & 1,000,000 & 12 E \\
2.44 & 732 & 1,800,000 & 1,000,000 & 6 E + 6 P \\
2.75 & 825 & 2,400,000 & 1,200,000 & 6 E + 6 P \\
2.97 & 891 & 3,000,000 & 1,200,000 & 6 E + 6 P \\
4.03 & 1,209 & 4,000,000 & 1,500,000 & 6 E + 6 P \\
\midrule
6 (F only) & 1,800 & 8,000,000 & 3,000,000 & 6 E + 6 P \\
8 (F only) & 2,400 & 8,000,000 & 3,000,000 & 6 E + 6 P \\
12 (F only) & 3,600 & 8,000,000 & 3,000,000 & 6 E + 6 P \\
\bottomrule
\end{tabular}
\end{table}

\begin{table}[!htbp]
\centering\small
\setlength{\tabcolsep}{4pt}
\renewcommand{\arraystretch}{1.15}
\caption{Numerical specifications for the mean-field and
continuum calculations. The stationary distributions are
obtained by self-consistent iteration of the zero-flux
recurrence on a truncated wealth domain $0\leq m\leq L$.
The reported tolerances refer to the $L^1$ difference
between successive iterations. All distributions are
computed from the model equations without loading
previously generated numerical data.}
\label{SI-tab:SI_MF}
\begin{tabular}{@{}p{0.16\linewidth} p{0.34\linewidth}
p{0.43\linewidth}@{}}
\toprule
Calculation & Control and method & Numerical settings \\
\midrule

3A MF curve &
Tanh bias, $\eta_\infty=1/4$; mean-wealth
bisection &
$L=70{,}000$; $L^1$ tolerance $10^{-11}$;
at most $1{,}200$ iterations per fixed-$p(0)$
solve \\

3A MC initializer &
Stationary MF solution at each prescribed
mean wealth; stratified sampling and
integer-total correction &
$L=70{,}000$; $L^1$ tolerance $10^{-11}$;
at most $1{,}200$ iterations per
fixed-$p(0)$ solve \\

3B MF distributions &
Tanh bias; prescribed
$\langle m\rangle=1.4,\ 2.6,\ 3.1$;
mean-wealth bisection &
$L=70{,}000$; $L^1$ tolerance $10^{-11}$;
at most $1{,}200$ iterations per
fixed-$p(0)$ solve \\

3B continuum &
Eq.~\ref{SI-distribution subcritical};
one horizontal matching shift
per distribution &
Minimize squared log-density residuals
over $20\leq m\leq200$ with
$p_{\mathrm{MF}}(m)\geq10^{-8}$;
no fitted vertical scale \\

3C near-critical &
Step, tanh, and Hill biases
($h=4,2$);
$\delta=10^{-5}$,
$p(0)=p_c+(1-p_c)\delta$ &
Fixed-$p(0)$ iteration at
$L=800{,}000$ and $1{,}600{,}000$;
$L^1$ tolerance $10^{-12}$;
at most $1{,}200$ iterations \\

4D MF reference &
Step bias, $\eta_\infty=1/4$;
near-critical solution at
$\delta=10^{-5}$ &
$L=2{,}000$; $L^1$ tolerance $10^{-9}$;
at most $600$ iterations;
survival function obtained by summation \\

\bottomrule
\end{tabular}
\end{table}

\begin{table}[!htbp]
\centering\small
\setlength{\tabcolsep}{5pt}\renewcommand{\arraystretch}{1.15}
\caption{Domain and iteration checks for Fig.~3C at $\delta=10^{-5}$.
Iteration counts are given for $L_1/L_2$.
The relative domain discrepancy is
$\epsilon_L =
\max_{1\le m\le20{,}000}
|p_{L_1}(m)-p_{L_2}(m)|/p_{L_2}(m)$,
where $L_1=800{,}000$ and $L_2=1{,}600{,}000$.
}
\label{SI-tab:SI_convergence}
\begin{tabular}{@{}l c c c c@{}}
\toprule
Interaction & Iterations & Final $L^1$ change & $\max_m|J_m|$ & $\epsilon_L$ \\
\midrule
Step & 92/92 & $9.36\times10^{-13}$ & $6.69\times10^{-13}$ & $4.490\times10^{-6}$ \\
Tanh & 93/93 & $9.70\times10^{-13}$ & $6.35\times10^{-13}$ & $4.491\times10^{-6}$ \\
Hill, $h=4$ & 104/104 & $8.74\times10^{-13}$ & $7.60\times10^{-13}$ & $4.492\times10^{-6}$ \\
Hill, $h=2$ & 103/103 & $9.93\times10^{-13}$ & $7.73\times10^{-13}$ & $4.510\times10^{-6}$ \\
\bottomrule
\end{tabular}
\end{table}

\end{document}